\documentclass[conference]{IEEEtran}
\IEEEoverridecommandlockouts
\usepackage{cite}
\usepackage{amsmath,amssymb,amsfonts}
\usepackage{algorithmic}
\usepackage{graphicx}
\usepackage{textcomp}
\usepackage{xcolor}
\usepackage[hyphens]{url}

\usepackage{algorithmic}
\usepackage[utf8]{inputenc}    % or appropriate for your template
\usepackage{amsmath}           % for \dfrac, \lfloor, etc.
\usepackage[ruled,vlined]{algorithm2e}
\usepackage{caption}
\usepackage{array}
\usepackage{orcidlink}
\usepackage[subrefformat=parens,labelformat=parens,font=small]{subcaption}
\usepackage{textcomp}
\usepackage{stfloats}
\usepackage{url}
\usepackage{verbatim}
\usepackage{graphicx}
\usepackage{booktabs}
\usepackage{bm}
\usepackage[table]{xcolor}  
\usepackage{booktabs} 
\usepackage{paralist}
\usepackage{multirow}
\usepackage{colortbl}
\usepackage{amsmath,amsfonts}% \mathbb, \Pr, \lceil, etc.
\usepackage[ruled,vlined]{algorithm2e}      % algorithm floats
\usepackage{tikz}
\usetikzlibrary{positioning}
\usepackage{tabularx,array}
\newcolumntype{C}{>{\centering\arraybackslash}X}
\AtBeginDocument{%
  \providecommand\BibTeX{{%
    Bib\TeX}}}

\def\BibTeX{{\rm B\kern-.05em{\sc i\kern-.025em b}\kern-.08em
T\kern-.1667em\lower.7ex\hbox{E}\kern-.125emX}}

\begin{document}
% Ensure letter paper
\pdfpagewidth=8.5in
\pdfpageheight=11in

%%%%%%%%%%%---SETME-----%%%%%%%%%%%%%
\newcommand{\iscasubmissionnumber}{465}
%%%%%%%%%%%%%%%%%%%%%%%%%%%%%%%%%%%%

\pagenumbering{arabic}

%\title{From Minutes to Seconds: Redefining the Five-Minute Rule for AI-Era Memory Hierarchies}
\title{Five-Minute Rule 40 Years Later: A First-Principles Revisit for Modern Memory Hierarchy}
\author{

\IEEEauthorblockN{Tong Zhang}
\IEEEauthorblockA{ScaleFlux, CA, USA}
\and
\IEEEauthorblockN{Vikram Sharma Mailthody}
\IEEEauthorblockA{NVIDIA, IL, USA}
\and
\IEEEauthorblockN{Fei Sun}
\IEEEauthorblockA{ScaleFlux, CA, USA}
\and
\IEEEauthorblockN{Linsen Ma}
\IEEEauthorblockA{ScaleFlux, CA, USA}
\and
\IEEEauthorblockN{Chris J. Newburn}
\IEEEauthorblockA{NVIDIA, IL, USA}

\and

\IEEEauthorblockN{Teresa Zhang}
\IEEEauthorblockA{Stanford University, CA, USA}
\and
\IEEEauthorblockN{Yang Liu}
\IEEEauthorblockA{ScaleFlux, CA, USA}
\and
\IEEEauthorblockN{Jiangpeng Li}
\IEEEauthorblockA{ScaleFlux, CA, USA}
\and
\IEEEauthorblockN{Hao Zhong}
\IEEEauthorblockA{ScaleFlux, CA, USA}
\and
\IEEEauthorblockN{Wen-Mei Hwu}
\IEEEauthorblockA{NVIDIA, IL, USA}

}
\maketitle
\pagestyle{empty}

\begin{abstract}
In 1987, Jim Gray and Gianfranco Putzolu introduced the five-minute rule, a simple, storage-memory-economics-based heuristic for deciding when data should live in DRAM rather than on storage. Subsequent revisits to the rule largely retained that economics-only view, leaving host costs, feasibility limits, and workload behavior out of scope. This paper revisits the rule from first principles, integrating host costs, DRAM bandwidth/capacity, and physics-grounded models of SSD performance and cost, and then embedding these elements in a constraint- and workload-aware framework that yields actionable provisioning guidance. We show that, for modern AI platforms, especially GPU-centric hosts paired with ultra-high-IOPS SSDs engineered for fine-grained random access, the DRAM$\leftrightarrow$flash caching threshold collapses from minutes to a few seconds. This shift reframes NAND flash memory as an \emph{active data tier} and exposes a broad research space across the hardware–software stack. We further introduce MQSim-Next, a calibrated SSD simulator that supports validation and sensitivity analysis and facilitates future architectural and system research. Finally, we present two concrete case studies that showcase the software system design space opened by such memory hierarchy paradigm shift. Overall, we turn a classical heuristic into an actionable, feasibility-aware analysis and provisioning framework and set the stage for further research on AI-era memory hierarchy.
\end{abstract}


%%%%%%%%%%%%%%%%%%%%%%%%%%%%%%%%%%%%%%%%%%%%%%%%%%%%%%%%%%%%%%
%%%%%%%%%%%%%%%%%%%%%%%%%%%%%%%%%%%%%%%%%%%%%%%%%%%%%%%%%%%%%%
\section{Introduction}
The evolution of storage hardware has long shaped data management system design. In the 1980s, databases operated on a two-tier hierarchy of DRAM and hard disk drives~(HDDs), when DRAM cost around \$120/KB and HDDs \$0.10/KB. This disparity led Jim Gray and Gianfranco Putzolu to ask: \textit{when is it more economical to keep data in memory rather than fetch it from disk?} Their 1987 \textit{five-minute rule}~\cite{gray19875} answered that 1KB records accessed more often than every five minutes should reside in DRAM, the point where DRAM ``rent'' is less than disk fetch cost. Later revisits in 1997~\cite{gray1997five}, 2007~\cite{graefe2007five}, and 2019~\cite{appuswamy2019five} updated the rule for advancing technology, with the latest still placing the DRAM-SSD threshold at the minute scale, echoing the adage: \emph{``Tape is dead, disk is tape, flash is disk.''} Yet these studies remained economics-only, overlooking host costs, feasibility limits, and workload behavior, offering little guidance for real-system provisioning.

As we fast forward to 2025, the storage landscape is undergoing another major shift. The rapid expansion of AI workloads is driving petabyte-scale working sets and highly diverse access patterns. This demand has fueled industry efforts such as NVIDIA's \textit{Storage-Next}\textsuperscript{TM}~\cite{qureshi2023gpu, ParkSamplingGNNGPU24, NVIDIA2025s73012}, which aims to unlock the full potential of NAND flash as a high-throughput, cost-effective extension of memory. In parallel, SSD vendors are investing heavily in developing {\bf Storage-Next SSDs} that deliver up to 10$\times$ higher IOPS per dollar, scaling efficiently as access granularity shrinks (e.g., 50M~IOPS at~512B vs. 10M~IOPS at~4KB)~\cite{nikkei-xtech-11065, digitimes2025samsung3dnand}. Complementing these advances, the HBF~(high-bandwidth flash) initiative~\cite{SanDisk2025_HBFKeynote} by SanDisk and SK~hynix targets 1TB/s per flash stack, signaling an industry trajectory toward NAND with bandwidth nearing that of HBM. Unlike prior NVM~(non-volatile memory) waves hindered by material and device limits, these developments build on mature NAND technology, making flash a plausible candidate for elevation from a capacity tier to an active tier of the memory hierarchy.

To reason about this trajectory, we revisit the five-minute rule from first principles. Our framework calibrates caching decisions with physics- and architecture-grounded inputs~(host costs and device behavior), incorporates feasibility constraints~(host IOPS and DRAM bandwidth/capacity), and embeds workload access intervals and service-level targets. This unified model (i) quantifies the impact of DRAM bandwidth, capacity, host IOPS, and SSD throughput on the DRAM-SSD caching threshold, (ii) translates them into concrete provisioning choices across platforms and workloads, and (iii) offers clear criteria for system feasibility and practical upgrade guidance. Under realistic architectural and device limits, we show that the DRAM-SSD caching threshold has {\bf collapsed from minutes to seconds}, redefining how memory and storage are provisioned for modern AI workloads. To support this framework, we develop \emph{MQSim-Next}, a calibrated SSD simulator built on MQSim~\cite{tavakkol2018mqsim, Mqsim-link} for validation, sensitivity studies, and future architectural research. Finally, we demonstrate its applicability through two case studies (i.e., key-value stores and approximate nearest neighbor search), highlighting the software design space enabled by this paradigm shift.

Together, these results argue for rethinking the memory hierarchy by elevating NAND flash from passive storage to an active tier, and provide architects with a practical toolkit for co-design across devices, hosts, and applications. Its major contributions are further summarized as follows:
\begin{compactitem}
\item A \textbf{first-principles reformulation} of the five-minute rule that integrates host costs, device behavior, and DRAM bandwidth/capacity. 
\item A \textbf{constraint-aware refinement} that bounds usable SSD IOPS via host capacity and tail-latency targets, replacing datasheet peaks with feasibility-aware IOPS. 
\item A \textbf{workload-aware platform framework} that combines access-interval profiles and service-level targets with system constraints to yield viability analysis and \emph{actionable} provisioning guidance. 
\item An \textbf{empirical finding} that GPU-centric hosts paired with Storage-Next SSDs can shrink the DRAM-flash caching threshold from minutes to seconds, together with guidance on when host-side limits dominate. 
\item \textbf{MQSim-Next}, a calibrated, physics-grounded SSD simulator used to validate model assumptions and support future architectural research in this space. 
\item Two \textbf{illustrative case studies} presented as initial steps for exploring the vast software/algorithm research space enabled by seconds-scale DRAM-flash caching.  
\end{compactitem}

This paper is organized as follows: Section~\ref{sec:background} reviews the background and states the research questions. Sections~\ref{sec:RQ1}–\ref{sec:RQ3} develop and validate the first-principles, workload-aware framework. Section~\ref{sec:mqsim} presents the MQSim-Next SSD simulator, and 
Section~\ref{sec:RQ4} presents the two case studies.

%%%%%%%%%%%%%%%%%%%%%%%%%%%%%%%%%%%%%%%%%%%%%%%%%%%%%%%%%%%%%%
%%%%%%%%%%%%%%%%%%%%%%%%%%%%%%%%%%%%%%%%%%%%%%%%%%%%%%%%%%%%%%
\section{Background and Motivation}
\label{sec:background}
The five-minute rule is a simple heuristic for data placement, yet it rarely informs real provisioning decisions. In its {\it economics-only} form, it overlooks host-side I/O costs and depends on vendor specifications.  Beyond economics, it ignores feasibility limits such as finite processor IOPS, latency and throughput targets, and DRAM bandwidth or capacity. We briefly revisit the classical rule, identify these omissions, and outline the research questions that motivate this work.

%%%%%%%%%%%%%%%%%%%%%%%%%%%%%%%%
\subsection{The Classical 5-Minute Rule and Its Limitation}
The rule makes a page-level decision: keep a page in DRAM when doing so is cheaper than fetching it from storage. This is expressed by the \emph{break-even interval}: if the expected inter-reference time is below this interval, cache the page; otherwise, leave it on storage. Balancing the ``rent'' on DRAM against the cost of repeated fetches yields:
\begin{equation}
\resizebox{\linewidth}{!}{$\tau_{\mathrm{break\text{-}even}}
= \left( \frac{\text{\# Pages per MB}}{\text{Storage Drive IOPS}} \right)
  \times \left( \frac{\text{Storage Drive Cost}}{\text{Cost of 1MB DRAM}} \right).
\label{eq:original_rule}$}\nonumber
\end{equation}
The classical rule can be written in a simpler, unit-consistent form. Let $C^{\text{page}}_{\text{DRAM}}$ denote the amortized capital cost of storing one page in DRAM, and $C^{\text{IO}}_{\text{SSD}}$ denote the amortized capital cost per storage access (i.e., SSD cost divided by its peak IOPS). Break-even occurs when keeping a page in DRAM over a reuse interval $T$ costs the same as repeatedly fetching it from storage:
\begin{equation}
T \cdot C^{\text{page}}_{\text{DRAM}} = C^{\text{IO}}_{\text{SSD}}.\nonumber
\end{equation}
Solving for $T$ yields:
\begin{equation}
T_{\text{break-even}} = \frac{C^{\text{IO}}_{\text{SSD}}}{C^{\text{page}}_{\text{DRAM}}}.\nonumber
\end{equation}
When host and bandwidth costs are ignored and full peak SSD IOPS is assumed, the calibrated formulation that will be presented later in Section~\ref{sec:RQ1} reduces to this classical expression. This expression makes explicit that the break-even interval is simply the ratio between per-access storage cost and per-page DRAM capital cost. Under HDD-era parameters in~\cite{gray19875}, this ratio yielded a break-even point in minutes, providing the historical context for our seconds-scale findings. However, this \textit{economics-only} view has the following key limits.

%\textcolor{blue}{Equivalently, break-even occurs when the amortized DRAM cost of retaining a page over a reuse interval equals the per-access cost of serving that page from storage. In other words, the product of reuse interval and DRAM cost per page balances the cost incurred per storage access. We express this relationship in unit-consistent form to make clear how time, capacity cost, and per-access cost interact in the classical rule}. Intuitively, the rule states that caching a page in DRAM is justified only when its rent is lower than the cost of fetching it from storage based on the device’s IOPS/\$ ratio. \textcolor{blue}{Under the original HDD-era assumptions in~\cite{gray19875}, this balance yielded a break-even interval on the order of minutes, providing the historical context for our seconds-scale findings}. However, this \textit{economics-only} view has the following key limits.

\noindent\textbf{(A) Insufficient realism.} The classical formulation treats host resources as free. In practice, issuing and completing I/O consumes CPU cycles, interrupts, and DRAM bandwidth, which are negligible for HDDs (100$\sim$200~IOPS) but significant for modern SSDs. Prior revisits~\cite{gray19875, gray1997five, graefe2007five, appuswamy2019five} also relied on vendor peak specs, overlooking architectural effects such as NAND physics, internal parallelism, and block-size scaling.\\[3pt]
\noindent\textbf{(B) Missing feasibility.} Optimizing only device prices (e.g., \$/GB or \$/IOPS) cannot ensure deployability. Real feasibility depends on host-side constraints~(e.g., submission/completion rate, latency and throughput targets, and DRAM bandwidth and capacity). Ignoring these factors can yield configurations unable to meet workload demands or service-level objectives.\\[3pt]
\noindent\textbf{Summary.} These gaps make the classical rule non-actionable for system design and provisioning. We therefore develop a feasibility-aware framework that models host resource usage and enforces practical system constraints, yielding accurate, actionable guidance. Unless stated otherwise, all cost terms denote amortized capital cost~(CapEx), consistent with Gray’s original rule.

%%%%%%%%%%%%%%%%%%%%%%%%%%%%%%%%
\subsection{Research Questions}
Our goal is to turn the classical rule from a heuristic into a basis for concrete and actionable framework, guided by the following three questions:\\[3pt]
\noindent\textbullet~\textbf{RQ1 (calibrated economics).} Retain the economic view but make it realistic by explicitly modeling host resource usage and first-principles SSD behavior. How does the break-even interval change under this calibrated model?\\[3pt]
\noindent\textbullet~\textbf{RQ2 (constraint-aware refinement).} Add feasibility constraints (i.e., processor I/O capacity and application latency targets). How do they reshape the break-even interval, and when do they become the primary constraint?\\[3pt]
\noindent\textbullet~\textbf{RQ3 (platform viability and guidance).} Integrate DRAM bandwidth/capacity limits and workload’s access-interval profile. Can a unified framework fusing \emph{economics}, \emph{workload}, and \emph{hardware constraints} assess viability and optimality, and, when needed, recommend upgrades?\\[3pt]
\indent Findings from these questions indicate that the DRAM-flash caching threshold has collapsed into the seconds regime due to the drastic elevation of IOPS/\$ of storage drives. As a result, the long-standing boundary between memory and storage has blurred, leading to the following research question:\\[3pt]
\noindent\textbullet~\textbf{RQ4 (software re-think).} As the DRAM-flash threshold drops to seconds, how should we rethink data-intensive software, and what principles should guide the redesign of data structures, access paths, and scheduling to fully exploit this new regime for throughput, efficiency, scalability, and cost?

Addressing these four questions shapes the remainder of this paper. It establishes a unified economics/feasibility framework with interpretable metrics for provisioning and upgrades, and it opens a principled design-space exploration under seconds-scale DRAM-flash caching.

%%%%%%%%%%%%%%%%%%%%%%%%%%%%%%%%%
\subsection{Discussion of Assumptions and Scope} 
All the modeling parameters in this study are derived from mature NAND flash technology and established roadmaps, in contrast to prior explorations that hypothesize active-memory roles for emerging NVMs. We model controller translation bandwidth and PCIe packet/bandwidth limits explicitly; in our evaluated configurations we provision these to be non-limiting, so the dominant bounds arise from NAND/channel physics and host capacity. Our goal is not to forecast product specifics, but to examine how feasibility and provisioning change once the full IOPS potential of NAND flash is unleashed. The framework is forward-looking yet physically grounded, and can be re-parameterized as devices and standards evolve. 

\noindent\textbf{Analytical vs.\ simulation components.} For clarity, we explicitly distinguish the analytical and simulation roles in this work. Sections~\ref{sec:RQ1}, \ref{sec:Constraint-aware}, and \ref{sec:RQ3} develop a closed-form, first-principles framework that derives break-even intervals, feasibility bounds, and platform-viability thresholds from device timing, host IOPS limits, DRAM bandwidth/capacity constraints, and workload access-interval profiles. Section~\ref{sec:mqsim} presents MQSim-Next, which models NAND timing, multi-plane concurrency, ECC behavior, and channel scheduling to characterize realistic device-level IOPS and latency trends. Section~\ref{sec:RQ4} then integrates these components: MQSim-Next provides calibrated device behavior, while the analytical framework determines usable IOPS, break-even thresholds, and system-level feasibility for the case-study evaluations.

%%%%%%%%%%%%%%%%%%%%%%%%%%%%%%%%
%%%%%%%%%%%%%%%%%%%%%%%%%%%%%%%%
\section{Calibrated Economic Model (RQ1)}\label{sec:RQ1}
This section grounds the break-even rule in a calibrated economics view: make host I/O costs explicit, and use architecture-derived SSD IOPS from a first-principles device model. We then present a quantitative case study showing that, under realistic configurations, GPUs paired with Storage-Next SSDs have shrunk the DRAM-flash break-even from minutes toward seconds; feasibility limits (processor IOPS and latency targets) are added in Section~\ref{sec:Constraint-aware}.

%This subsection introduces the calibrated economic view that underpins our framework. We first define a simple host and device model and restate the classical, economic-only break-even interval using realistic inputs: explicit I/O-induced host resource usage cost, and a storage performance term given by architecture-derived peak IOPS computed from a first-principles SSD model rather than datasheet claims. We then elaborate on deriving the SSD-related metrics via device-level SSD modeling. This subsection concludes with a quantitative case study showing that, under this calibrated formulation, a GPU host paired with Storage-Next SSDs can shorten the break-even interval from minutes toward seconds. Feasibility constraints such as processor I/O budgets and latency objectives are introduced in Section~\ref{sec:Constraint-aware}.

%%%%%%%%%%%%%%%%%%%%%%%%%%%%%%%%%%%%%%%%%%%%%%%%%%%%%%%%%%%%%%
\subsection{System Model and Calibrated Economic Break-even}\label{sec:systemmodel}
\begin{figure}[htbp]
    \centering
\includegraphics[width=\linewidth]{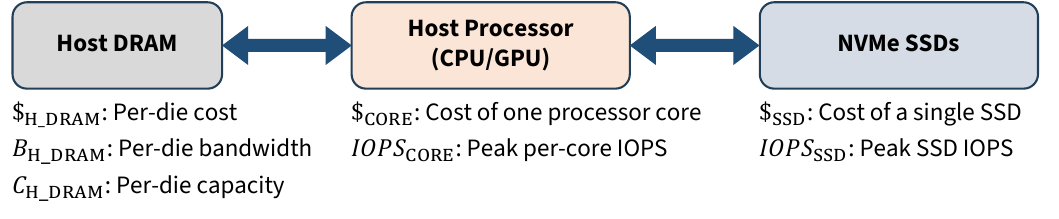}
    \caption{Simplified system architecture used to derive the new break-even interval formulation.}
    \label{fig:highlevel}
\end{figure}

Fig.~\ref{fig:highlevel} presents a first-order host–device view of the I/O path: a host processor (CPU or GPU), a directly attached, multi-channel DRAM subsystem, and NVMe SSDs. We assume an optimal zero-copy read path~\cite{Khalidi-zero-copy-95} to minimize host DRAM bandwidth usage by avoiding extra kernel$\leftrightarrow$user copies. Now consider an $l_{\text{blk}}$-byte block accessed periodically with reuse interval $\tau_{\text{intvl}}$. Absent caching in host DRAM, the system repeatedly retrieves this block from the SSD, incurring cumulative cost across the following three components:\\[3pt] 

\begin{comment}
    
\begin{table}[htbp]
\centering
\caption{System-level parameters used in the break-even interval formulation.}
\label{tab:sys-params}
\renewcommand{\arraystretch}{1.2}
\rowcolors{2}{gray!10}{white}
\begin{tabular}{m{1.6cm} m{1.9cm} m{3.8cm}}
\toprule
\textbf{Symbol} & \textbf{Component} & \textbf{Definition} \\
\midrule
$ l_{\text{blk}} $ & I/O access & Data access block size \\
$ \$_{\text{H\_DRAM}} $ & Host DRAM & Cost of one DRAM die \\
$ B_{\text{H\_DRAM}} $ & Host DRAM & Bandwidth of one DRAM die \\
$ C_{\text{H\_DRAM}} $ & Host DRAM & Capacity of a single DRAM die \\
$ \$_{\text{CORE}} $ & Host processor & Cost of one processor core \\
$ IOPS_{\text{CORE}} $ & Host processor & Maximum IOPS supported by one processor core \\
$ \$_{\text{SSD}} $ & SSD & Cost of a single SSD \\
$ IOPS_{\text{SSD}} $ & SSD & Maximum SSD IOPS for the access block size $ l_{\text{blk}} $\\
\bottomrule
\end{tabular}
\end{table}
\end{comment}
\noindent\textbf{Host processor cost}: Each I/O involves driver-level work such as queue management and interrupt handling. Given the per-core IOPS capacity of \( IOPS_{\text{CORE}} \)\footnote{For simplicity, we assume \(IOPS_{\text{CORE}}\) is independent of the workload's read-to-write ratio; processing read and write requests incurs similar processor overhead.} and per-core cost \( \$_{\text{CORE}} \), we can express the cost on host processor as $\frac{\$_{\text{CORE}}}{IOPS_{\text{CORE}}}\cdot \frac{1}{\tau_{\text{intvl}}}$.\\[3pt]
\noindent\textbf{Host DRAM bandwidth cost:} Each I/O transfers \(l_{\text{blk}}\) bytes into host DRAM, consuming the DRAM bandwidth. We model its cost as $\frac{l_{\text{blk}}\,\$_{\text{H\_DRAM}}}{B_{\text{H\_DRAM}}}\cdot \frac{1}{\tau_{\text{intvl}}}$,
which is appropriate for bandwidth-bound systems common in modern AI infrastructure. If bandwidth is ample and capacity is the constraint, a capacity-based DRAM cost model would be more appropriate.\\[3pt]
\noindent \textbf{SSD access cost}: Given its cost of $\$_{\text{SSD}}$ and peak IOPS of \( IOPS_{\text{SSD}} \), the access cost on SSD is $\frac{\$_{\text{SSD}}}{IOPS_{\text{SSD}}}\cdot \frac{1}{\tau_{\text{intvl}}}$.\footnote{Consistent with the classical economic-only view, we assume the host can fully utilize the SSD’s peak random IOPS for given data access block size and read-to-write ratio. Later sections incorporate hardware and workload constraints that bound usable IOPS and can change the effective cost.}\\[3pt]
Therefore, if we cache a data block being accessed with an interval of \( \tau_{\text{intvl}} \) in DRAM, we can express the saved cost as:
\begin{equation}
\resizebox{0.92\linewidth}{!}{$
\$_{\text{saving}} =
\left(
\frac{\$_{\text{CORE}}}{IOPS_{\text{CORE}}}
+ \frac{l_{\text{blk}} \cdot \$_{\text{H\_DRAM}}}{B_{\text{H\_DRAM}}}
+ \frac{\$_{\text{SSD}}}{IOPS_{\text{SSD}}}
\right) \cdot \frac{1}{\tau_{\text{intvl}}}\ .
\label{eq:saving} $}\nonumber
\end{equation}

\noindent
By caching the block in host DRAM, the system avoids this recurring cost. However, doing so requires reserving a portion of host DRAM capacity over time, which incurs a ``rent'':
\begin{equation}
\$_{\text{rent}} = \frac{l_{\text{blk}}}{C_{\text{H\_DRAM}}} \cdot \$_{\text{H\_DRAM}}.
\label{eq:rent}\nonumber
\end{equation}
The break-even interval \( \tau_{\text{break-even}} \) is the access interval at which the memory rent equals the cost saved by avoiding repeated I/O operations. Solving \( \$_{\text{rent}} = \$_{\text{saving}} \) yields:
\begin{equation}\label{eq:breakeven}
\resizebox{0.91\linewidth}{!}{$
  \tau_{\text{break-even}} =
  \left(
    \frac{\$_{\text{CORE}}}{IOPS_{\text{CORE}}} +
    \frac{l_{\text{blk}}\cdot\$_{\text{H\_DRAM}}}{B_{\text{H\_DRAM}}} +
    \frac{\$_{\text{SSD}}}{IOPS_{\text{SSD}}}
  \right)
  \cdot \frac{C_{\text{H\_DRAM}}}{l_{\text{blk}}\cdot\$_{\text{H\_DRAM}}}\,.
$}
\end{equation}
We note that the numerator has units of \$/IO, while the denominator represents \$/MB amortized over capacity, so $\tau_{\text{break-even}}$ has units of time, as expected. This calibrated formulation preserves Gray's intuition: balance the DRAM ``rent'' against the cost of serving accesses from storage. It (i) explicitly charges I/O-induced host resources, and (ii) replaces datasheet peaks with SSD performance and cost derived from device behavior. Thus, it offers a more accurate economic criterion for deciding when DRAM caching is justified. In this model, \(IOPS_{\text{SSD}}\) and \(\$_{\text{SSD}}\) are parameters rather than constants; Section~\ref{sec:devicemodeling} derives them from first principles using a device-level SSD model.

%%%%%%%%%%%%%%%%%%%%%%%%%%%%%%%%%%%%%%%%%%%%%%%%%%%%%%%%%%%%%%
\subsection{First-Principles SSD Modeling}
\label{sec:devicemodeling}
\begin{figure}[htbp]
    \centering
\includegraphics[width=\linewidth]{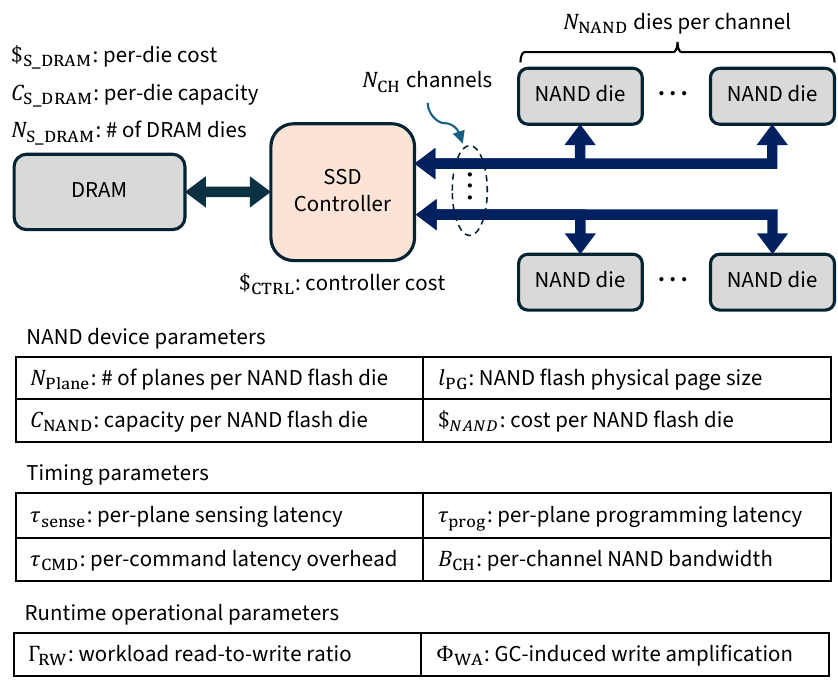}
    \caption{SSD architecture with key parameters for modeling performance and cost.}
    \label{fig:SSDmodel}
\end{figure}

We adopt a first-principles model of SSD performance and cost grounded in NAND device architecture. As illustrated in Fig.~\ref{fig:SSDmodel}, an SSD consists of a controller, SSD-internal DRAM for the FTL~(flash translation layer), and a NAND subsystem. The channel command time $\tau_{\text{CMD}}$ denotes bus occupancy per command; in conventional NAND with an 8-bit shared command/data bus, $\tau_{\text{CMD}}\!\approx\!1.2\,\mu\text{s}$~\cite{ONFI-Specs-Page}, whereas modern devices employ the SCA I/O protocol to reduce it to 100-200ns~\cite{SCA}, improving effective bandwidth. We model performance from sensing, programming, and command latencies, and omit erase operations since each erase clears megabytes of data and contributes negligibly in steady state. For broader background on SSD and NAND flash technology, see~\cite{micheloni2013inside, compagnoni2017reviewing}.

In our first-principles model, peak SSD performance is bounded by NAND parallelism, channel bandwidth, controller address translation bandwidth, and PCIe packet rate limits. For tractability, all host-issued requests use the same block size $l_{\text{blk}}$. Let $IOPS_{\text{NAND}}^{(\text{peak})}$ denote the maximum IOPS deliverable by a single NAND die, and $IOPS_{\text{CH}}^{(\text{peak})}$ the maximum IOPS sustainable by a single channel. With $N_{\text{CH}}$ channels and $N_{\text{NAND}}$ dies per channel, we can formulate the memory-device-limited IOPS $IOPS_{\mathrm{dev}}^{(\mathrm{peak)}}$ as
\begin{equation}
\resizebox{\linewidth}{!}{$
    IOPS_{\mathrm{dev}}^{(\mathrm{peak)}} =\frac{\Gamma_{\text{RW}}+1}{\Gamma_{\text{RW}}+2\Phi_{\text{WA}}-1}
    \cdot N_{\text{CH}} \cdot \min\!\left( N_{\text{NAND}} \cdot IOPS_{\text{NAND}}^{\text{(peak)}},\; IOPS_{\text{CH}}^{\text{(peak)}}\right),$
    }\nonumber
\end{equation}
where \(\Gamma_{\text{RW}}\) denote the read-to-write ratio and \(\Phi_{\text{WA}} \ge 1\) captures write-amplification caused by background garbage collection~(GC). We further define $IOPS_{\text{xlat}}^{(\text{peak})}$ as the maximum IOPS supported by controller address translation bandwidth, and $IOPS_{\text{pcie}}^{(\text{peak})}$ as the maximum IOPS supported by the PCIe packet rate. Accordingly, 
the overall peak SSD IOPS is
\begin{equation}
    \label{eq:SSDIOPSpeak} \resizebox{0.91\linewidth}{!}{$ IOPS_{\mathrm{SSD}}^{(\mathrm{peak})}
=
\min\!\left(
IOPS_{\mathrm{dev}}^{(\mathrm{peak})},
IOPS_{\mathrm{xlat}}^{(\mathrm{peak})},
IOPS_{\mathrm{pcie}}^{(\mathrm{peak})}
\right).$}
\end{equation}

\noindent\underline{Formulation of $IOPS_{\text{NAND}}^{\text{(peak)}}$}: Since a physical page must be programmed as a unit, the controller coalesces host random writes into full-page sequential writes. Thus, within one program interval \(\tau_{\text{prog}}\), a die can commit \(N_{\text{Plane}}\cdot l_{\text{PG}}/l_{\text{blk}}\) blocks. For reads, within one sense interval \(\tau_{\text{sense}}\), a die can fetch \(N_{\text{Plane}} \) blocks. Combining the workload read-to-write ratio \(\Gamma_{\text{RW}}\) and intra-SSD write amplification \(\Phi_{\text{WA}}\), we have that, the read fraction is \( R_r=(\Gamma_{\text{RW}}+\Phi_{\text{WA}}-1)/(\Gamma_{\text{RW}}+2\Phi_{\text{WA}}-1)\) and the write fraction is \( R_w=\Phi_{\text{WA}}/(\Gamma_{\text{RW}}+2\Phi_{\text{WA}}-1)\). Hence, the per-die peak IOPS is
\begin{equation}\label{eq:IOPSNAND}
    IOPS_{\text{NAND}}^{\text{(peak)}}
    = R_r\cdot \frac{N_{\text{Plane}}}{\tau_{\text{sense}}}
      \;+\;
      R_w\cdot \frac{N_{\text{Plane}}\cdot l_{\text{PG}}}{\tau_{\text{prog}}\cdot l_{\text{blk}}}\,.\nonumber
\end{equation}

\noindent\underline{Formulation of $IOPS_{\text{CH}}$}: With channel bandwidth \(B_{\text{CH}}\), reading a size-\(l_{\text{blk}}\) block occupies the channel for
\(\tau_{\mathrm{R}}=\tau_{\mathrm{CMD}}+l_{\text{blk}}/B_{\text{CH}}\), so one channel can deliver up to \(1/\tau_{\mathrm{R}}\) read IOPS. A program transfers a full physical page of size \(l_{\text{PG}}\), occupying the channel for \(\tau_{\mathrm{W}}=\tau_{\mathrm{CMD}}+l_{\text{PG}}/B_{\text{CH}}\). Each program commits \(l_{\text{PG}}/l_{\text{blk}}\) blocks, so each channel can support up to $l_{\text{PG}}/(l_{\text{blk}}\cdot \tau_{\text{W}})$ write IOPS.
Hence, the peak IOPS sustainable by each channel is
\begin{equation}\label{eq:IOPSCH}
  IOPS_{\text{CH}}^{\text{(peak)}}
  = R_r\cdot \frac{1}{\tau_{\mathrm{CMD}} + \frac{l_{\text{blk}}}{B_{\text{CH}}}}
    \;+\;
    R_w\cdot
    \frac{1}{\frac{l_{\text{blk}}}{l_{\text{PG}}}\,\tau_{\mathrm{CMD}} + \frac{l_{\text{blk}}}{B_{\text{CH}}}}\,.\nonumber
\end{equation}

\noindent\underline{Formulation of $IOPS_{\mathrm{xlat}}^{(\mathrm{peak})}$}: Each random request requires a logical-to-physical address translation in the FTL. If each FTL entry is $b_{\mathrm{FTL}}$ bytes and $B_{\mathrm{SSD\_DRAM}}$ denotes the bandwidth of the SSD-internal DRAM storing FTL metadata, then under the conservative assumption of no translation-cache hits, the translation-bandwidth-limited peak IOPS is 
\begin{equation}
IOPS_{\mathrm{xlat}}^{(\mathrm{peak})}
=
\frac{B_{\mathrm{SSD\_DRAM}}}{b_{\mathrm{FTL}}}. \nonumber
\end{equation}
With $b_{\mathrm{FTL}}=8$B per mapping entry and a controller DRAM bandwidth of $B_{\mathrm{SSD\_DRAM}}=40$GB/s, we obtain $IOPS^{(\mathrm{peak})}_{\mathrm{xlat}} \approx 5$G IOPS, well above the NAND/channel-limited peak in our evaluated Storage-Next configurations~(tens of millions of IOPS).

\noindent\underline{Formulation of $IOPS_{\mathrm{pcie}}^{(\mathrm{peak})}$}: At small block sizes, the PCIe link may be limited either by aggregate bandwidth or by packet-processing rate. Let $B_{\mathrm{PCIe}}$ denote the effective PCIe bandwidth, $PPS_{\mathrm{host}}$ the maximum packet rate supported by the PCIe root complex, and $n_{\mathrm{pkt}}(l_{\mathrm{blk}})$ the number of PCIe transactions required to serve a request of size $l_{\mathrm{blk}}$. The interconnect-limited peak IOPS is therefore
\begin{equation}
IOPS_{\mathrm{pcie}}^{(\mathrm{peak})}
=
\min\!\left(
\frac{B_{\mathrm{PCIe}}}{l_{\mathrm{blk}}},
\frac{PPS_{\mathrm{host}}}{n_{\mathrm{pkt}}(l_{\mathrm{blk}})}
\right).
\end{equation}
For a representative PCIe Gen7 x4 link with nominal bandwidth $B_{\mathrm{PCIe}} \approx 64$GB/s, the first bandwidth-limited term is over 120M,  well above the 50M-class NAND/channel peak studied in this work. For the second packet-rate term, sustaining 50M IOPS requires
\(
PPS_{\mathrm{host}}
\ge
50\text{M} \cdot n_{\mathrm{pkt}}(512\text{B}).
\) 
In our evaluated configurations that have fine-grained I/Os, the NAND channel I/O rate limitations are significantly more stringent than the bandwidth of the device's interface to the PCIe.
%%Even if each I/O generates on the order of $4$–$8$ PCIe/NVMe transactions, this corresponds to 200–400M transactions/s. In our evaluated configurations, PCIe is provisioned so that this requirement does not dominate the NAND/channel-limited peak; the packet-rate term is included to make this feasibility constraint explicit.

As shown in Fig.~\ref{fig:SSDmodel}, the total SSD cost aggregates the controller, all the NAND dies, and the SSD-internal DRAM used primarily for FTL:
\begin{equation}
  \$_{\text{SSD}}
  = \$_{\text{CTRL}}
    + N_{\text{CH}}\cdot N_{\text{NAND}}\cdot \$_{\text{NAND}}
    + N_{\text{S\_DRAM}}\cdot \$_{\text{S\_DRAM}}\, .\nonumber
\end{equation}
Here \(N_{\text{S\_DRAM}}\) is the number of SSD-internal DRAM dies. If each FTL entry is \(b_{\text{FTL}}\) bytes (e.g., 4-8 bytes) and the minimum access granularity is 512B, the maximum FTL size is
\begin{equation}
  C_{\text{FTL}}
  = \frac{N_{\text{CH}}\cdot N_{\text{NAND}}\cdot C_{\text{NAND}}}{512\mathrm{B}}
    \cdot b_{\text{FTL}}\, .\nonumber
\end{equation}
Given the capacity per DRAM die \(C_{\text{S\_DRAM}}\), the required SSD-internal DRAM die count is
\begin{equation}
  N_{\text{S\_DRAM}}
  = \left\lceil \frac{C_{\text{FTL}}}{C_{\text{S\_DRAM}}} \right\rceil
  = \left\lceil
      \frac{N_{\text{CH}}\cdot N_{\text{NAND}}\cdot C_{\text{NAND}}\cdot b_{\text{FTL}}}
           {512\mathrm{B}\cdot C_{\text{S\_DRAM}}}
    \right\rceil .\nonumber
\end{equation}

These formulations make explicit how SSD IOPS and cost scale with architectural choices and operating parameters, and they expose the coupling between performance and capacity. Unlike vendor specifications that reflect a few fixed configurations, this first-principles model supports architecture-aware reasoning across a broad design space and workload settings.

%%%%%%%%%%%%%%%%%%%%%%%%%%%%%%%%%%%%%%%%%%%%%%%%%%%%%%%%%%%%%%
\subsection{Quantitative Study} \label{sec:casestudy1}
To clarify the origin of the ``Storage-Next'' IOPS values used throughout this paper, we emphasize that they are not vendor-projected datasheet numbers. Instead, they are derived directly from the formulations presented in Sec.~\ref{sec:devicemodeling}, using representative NAND timing and architectural parameters summarized in Table~\ref{tab:case-ssd} for three NAND types: (i) SLC (1bit/cell) devices optimized for low latency and high IOPS (e.g., Kioxia XL-Flash~\cite{shiozawa2020emerging} and Samsung Z-NAND~\cite{cheong2018flash}); (ii) TLC~(3bits/cell) operated in pseudo-SLC~(pSLC) mode; and (iii) standard TLC. For example, under the SLC configuration in Table~\ref{tab:case-ssd}~($\tau_{\text{sense}}=5\,\mu$s, $\tau_{\text{prog}}=50\,\mu$s, $N_{\text{Plane}}=6$, $N_{\text{CH}}=20$, $N_{\text{NAND}}=4$, $B_{\text{CH}}=3.6$\,GB/s, $\tau_{\text{CMD}}=150$\,ns), and assuming $\Gamma_{\text{RW}}=90{:}10$ and $\Phi_{\text{WA}}=3$, the model yields $\mathrm{IOPS}^{(\mathrm{peak})}_{\mathrm{SSD}}\approx 57$M at 512B and $\mathrm{IOPS}^{(\mathrm{peak})}_{\mathrm{SSD}}\approx 11$M at 4KB. The high small-block IOPS arises from the combination of short sensing latency, intra-die multi-plane parallelism, and channel bandwidth that scales approximately as $B_{\text{CH}}/l_{\text{blk}}$ for small blocks when $\tau_{\text{CMD}}$ is reduced via SCA. Therefore, the 50M-class IOPS regime at 512B used in our Storage-Next configuration reflects a direct first-principles evaluation of NAND timing and architectural parallelism, rather than an assumed external projection.

\begin{figure}[htbp]
    \centering
\includegraphics[width=\linewidth]{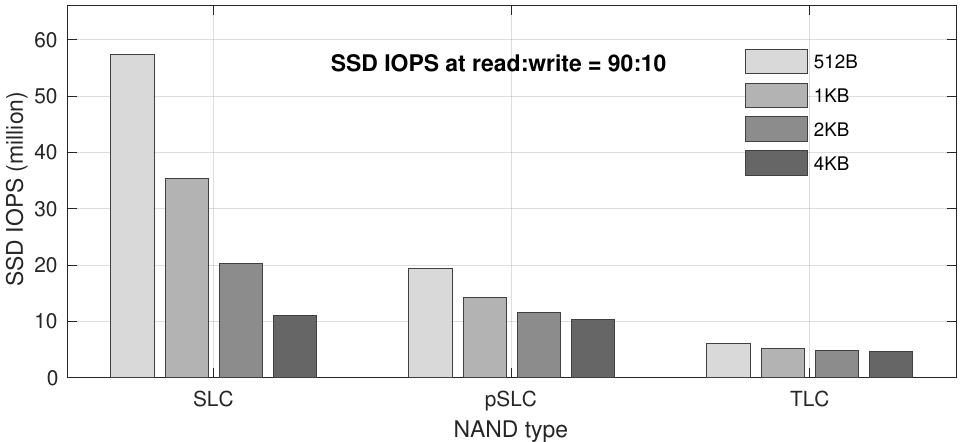}
    \caption{Storage-Next SSD peak IOPS under different configurations and workload read-to-write ratio of 90:10.}
    \label{fig:SSDIOPS}
\end{figure}

The sensing and programming latencies and the multi-plane organization parameters in Table~\ref{tab:case-ssd}~(e.g., $\tau_{\mathrm{sense}}$, $\tau_{\mathrm{prog}}$, and $N_{\mathrm{Plane}}$) are representative of published 3D-NAND device characterizations and
independent multi-plane implementations~(e.g.,~\cite{cheong2018flash, khakifirooz2021nand, pekny2022independent}), and reflect contemporary low-latency NAND designs. We demonstrate the framework with a quantitative study of break-even intervals across realistic system configurations. In this study, we fix \(\Gamma_{\text{RW}}=90{:}10\), reflecting read-heavy AI workloads, and conservatively set \(\Phi_{\text{WA}}=3\). Fig.~\ref{fig:SSDIOPS} shows peak IOPS for SLC, pSLC, and TLC across 512B-4KB blocks. As Eq.~\ref{eq:SSDIOPSpeak} indicates, overall SSD IOPS is bounded by the smaller of the device limit \(IOPS_{\text{NAND}}^{\text{(peak)}}\) and the channel limit \(IOPS_{\text{CH}}^{\text{(peak)}}\). The device term depends mainly on \(\tau_{\text{sense}}\) and \(\tau_{\text{prog}}\) and varies only weakly with \(l_{\text{blk}}\); the channel term depends strongly on \(l_{\text{blk}}\) and, with small \(\tau_{\text{CMD}}\), scales roughly as \(B_{\text{CH}}/l_{\text{blk}}\). For TLC, long \(\tau_{\text{sense}}\) and \(\tau_{\text{prog}}\) keep \(IOPS_{\text{NAND}}^{\text{(peak)}}\) low, so the device side limits IOPS for all \(l_{\text{blk}}\), producing only slight variation with block size. For SLC, very short \(\tau_{\text{sense}}\) and \(\tau_{\text{prog}}\) raise \(IOPS_{\text{NAND}}^{\text{(peak)}}\); small blocks are device-limited, while larger blocks become channel-limited, yielding a strong, though not perfectly proportional, increase of IOPS as \(l_{\text{blk}}\) decreases. pSLC falls between SLC and TLC across all sizes. Storage-Next SSDs are designed to exploit this regime: they provide scalable small-block IOPS, especially with SLC or pSLC, whereas conventional SSDs remain nearly flat at $\le$4KB due to 4KB-oriented ECC/controller architecture.
\begin{table}[t]
  \centering
  \caption{Key SSD parameters. SLC timing values are representative of low-latency NAND devices  (e.g., XL-Flash~\cite{shiozawa2020emerging}, Z-NAND~\cite{cheong2018flash}).  $\tau_{\mathrm{CMD}}$ reflects SCA protocol timing~\cite{SCA}. $l_{\mathrm{PG}}$ and interface bandwidth follow ONFI specifications~\cite{ONFI-Specs-Page}}.\vspace*{-6pt}\label{tab:keySSDpara}
  \setlength{\tabcolsep}{4pt}
  \renewcommand{\arraystretch}{1.1}
\label{tab:case-ssd}
  % --- top table ---
  \begin{tabularx}{\linewidth}{|C|C|C|C|C|C|}
    \hline
      & $\tau_{\text{sense}}$ & $\tau_{\text{prog}}$ & $l_{\text{PG}}$ & $N_{\text{Plane}}$ & $C_{\text{NAND}}$ \\
    \hline
      SLC  & $5\,\mu$s  & $50\,\mu$s  & 4KB  & 6 & 32GB  \\ \hline
      pSLC & $20\,\mu$s & $150\,\mu$s & 16KB & 4 & 42GB  \\ \hline
      TLC  & $40\,\mu$s & 1ms         & 16KB & 4 & 128GB \\ \hline
  \end{tabularx}

  \vspace{4pt}

  % --- bottom table ---
  \begin{tabularx}{\linewidth}{|C|C|C|C|C|}
    \hline
      $\tau_{\text{CMD}}$ & $B_{\text{CH}}$ & $N_{\text{CH}}$ & $N_{\text{NAND}}$ & $C_{\text{S\_DRAM}}$ \\
    \hline
      150ns & 3.6GB/s & 20 & 4 & 3GB \\ \hline
  \end{tabularx}
\end{table}

To further demonstrate that the ``50M-class'' small-block regime is not tied to a single parameter point, we perform a brief sensitivity sweep over three primary architectural knobs in Eq.~\ref{eq:SSDIOPSpeak}: the number of channels ($N_{\mathrm{CH}}$), dies per channel ($N_{\mathrm{NAND}}$), and per-command overhead ($\tau_{\mathrm{CMD}}$). Table~\ref{tab:sensitivity} reports the resulting peak IOPS for the SLC configuration under the same workload parameters as Fig.~\ref{fig:SSDIOPS} ($\Gamma_{\mathrm{RW}}{=}90{:}10$, $\Phi_{\mathrm{WA}}{=}3$). Across pessimistic-to-optimistic settings, 512B peak IOPS remains in the tens-of-millions range, indicating that the key small-block scaling trend (and the resulting seconds-scale implications studied in later sections)
is robust to moderate variation of these design parameters. Because the break-even expressions depend primarily on IOPS/\$ rather than any single knob, these variations shift absolute thresholds but do not revert the qualitative finding that modern small-block IOPS can push the DRAM$\leftrightarrow$flash boundary into the seconds regime.

\begin{table}[h]
  \centering
  \caption{Sensitivity of peak SSD IOPS (SLC) to architectural scaling knobs.
  All other parameters follow Table~\ref{tab:case-ssd}.}
  \label{tab:sensitivity}
  \vspace{-0.5em}
  \renewcommand{\arraystretch}{1.2}
\resizebox{\linewidth}{!}{%
  \begin{tabular}{lccc|cc}
    \toprule
    Setting & $N_{\mathrm{CH}}$ & $N_{\mathrm{NAND}}$ & $\tau_{\mathrm{CMD}}$ &
    IOPS@512B & IOPS@4KB \\
    \midrule
    Pessimistic & 16 & 3 & 200ns & 39.4M & 8.5M \\
    Baseline (Table~\ref{tab:case-ssd}) & 20 & 4 & 150ns & 57.4M & 11.1M \\
    Optimistic & 24 & 5 & 100ns & 79.3M & 13.8M \\
    \bottomrule
  \end{tabular}}
  \vspace{-1.0em}
\end{table}

Since Eq.~\ref{eq:breakeven} includes costs in both numerator and denominator, we normalize all components to the NAND-die cost for fair comparison. Let $\alpha_{\text{CTRL}}$, $\alpha_{\text{S\_DRAM}}$, $\alpha_{\text{H\_DRAM}}$, and $\alpha_{\text{CORE}}$ denote normalized costs of the SSD controller, SSD-internal DRAM, host DRAM, and host cores. Table~\ref{tab:normalized_costs} lists values for CPU+DDR and GPU+GDDR platforms. All numbers derive from manufacturing parameters (e.g., die area and process node), rather than market price, avoiding buyer bias. DDR and NAND have comparable die areas, so DDR’s cost is set to~1, GDDR to~2 due to higher pin counts and tighter thermal limits. Based on internal design data, $\alpha_{\text{CTRL}}=15$ (reflecting controller complexity on 12–7 nm nodes). A server-class CPU core has cost~4 with 1M IOPS/core, while a GPU SM has cost~3 with 4M IOPS/SM, following NVIDIA’s SCADA~(SCaled Accelerated Data Access) platform~\cite{newburn_scada_2024} on the NVIDIA Hopper generation of GPUs. Though actual prices vary, this normalized model provides a consistent, architecture-based basis for comparison.
\begin{table}[htbp]
\centering
\caption{Normalized cost and performance parameters under different compute+memory configurations.}
\label{tab:normalized_costs}
\renewcommand{\arraystretch}{1.2}
\resizebox{\linewidth}{!}{%
\begin{tabular}{lcccccccc}
\toprule
\textbf{Platform} & \( \alpha_{\text{H\_DRAM}} \) & \( B_{\text{H\_DRAM}} \) & \( C_{\text{H\_DRAM}} \) & \( \alpha_{\text{CORE}} \) & \( \text{IOPS}_{\text{CORE}} \) & \( \alpha_{\text{CTRL}} \) & \( \alpha_{\text{S\_DRAM}} \) \\
\midrule
CPU+DDR & 1 & 3\,GB/s & 3\,GB & 4 & 1M & 15 & 1 \\
GPU+GDDR & 2 & 80\,GB/s & 2\,GB & 3 & 4M & 15 & 1 \\
%GPU+HBM & 100 & 2\,TB/s & 64\,GB & 3 & 4M & 15 & 1 \\
\bottomrule
\end{tabular}%
}
\end{table}

\begin{figure}[htbp]
    \centering
    \includegraphics[width=\linewidth]{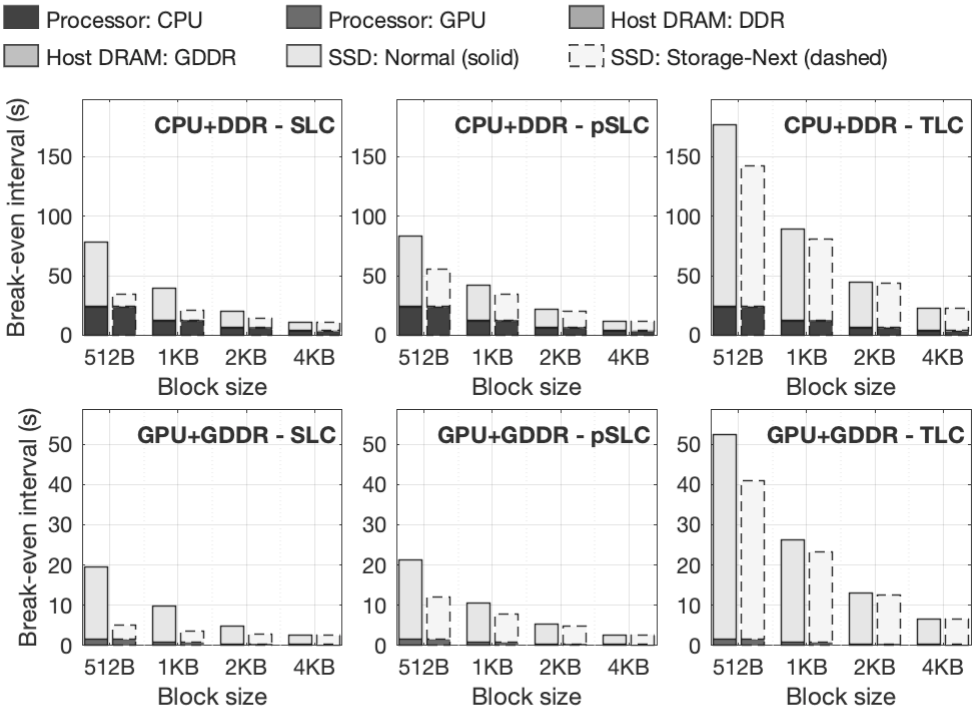}
    \caption{Break-even interval across configurations. Each stack shows contributions from host processor, DRAM, and SSD.}
    \label{fig:breakevenexample}
\end{figure}

We then compute the break-even intervals shown in Fig.~\ref{fig:breakevenexample}. For each block size, the left bar represents the \emph{Normal-SSD} baseline (flat IOPS for $\le$4\,KB), and the right bar the \emph{Storage-Next SSD} whose IOPS increases as block size shrinks. Following Gray’s formulation, we assume full utilization of peak SSD IOPS, i.e., $IOPS_{\text{SSD}}=IOPS_{\text{SSD}}^{\text{(peak)}}$ in Eq.~\ref{eq:breakeven}. Each stacked bar decomposes the interval into processor, DRAM, and SSD components, revealing how architectural and device parameters shape placement decisions. As NAND sensing latency grows from 5$\mu$s (SLC) to 40$\mu$s (TLC), SSD IOPS/\$ drops and its share in total cost rises. Larger block sizes yield shorter intervals due to higher DRAM ``rent'': under SLC on CPU+DDR, the interval decreases from $\sim$34s at 512B to $\sim$10s at 4KB. GPU platforms show much shorter intervals; for SLC at 512\,B, the break-even time falls from $\sim$34s (CPU+DDR) to $\sim$5s (GPU+GDDR), a 7$\times$ reduction. Across all block sizes, Storage-Next SSDs consistently outperform Normal-SSDs for sub-4\,KB requests, with the largest gaps in SLC devices where small-block IOPS scaling dominates.  These results confirm that a first-principles framework, grounded in device physics rather than vendor specs, better captures cost–performance trade-offs. Storage-Next SSDs with scalable IOPS cut the break-even interval from minutes to seconds, reaching single digits on GPU platforms, and elevate NAND flash to an active tier of the memory hierarchy. All timing and bandwidth parameters used in this section are grounded in  published device characterizations and interface specifications, and can be re-parameterized as technologies evolve. The SSD-only component shown in Fig.~\ref{fig:breakevenexample} corresponds to the classical Gray analysis but with drastically different cost parameters, such that the effective break-even threshold drops dramatically.

%%%%%%%%%%%%%%%%%%%%%%%%%%%%%%%%
\section{Constraint-aware Break-even (RQ2)} \label{sec:Constraint-aware}
The calibrated economic model above assumes, as in the original 5-minute rule, that system fully utilizes SSD’s peak IOPS. Here, we relax this assumption and make the break-even analysis feasibility-aware by introducing two practical constraints that bound usable SSD IOPS: (i) application-level read latency and (ii) the platform’s total host IOPS capacity.

\begin{figure}[htbp]
    \centering
    \includegraphics[width=\linewidth]{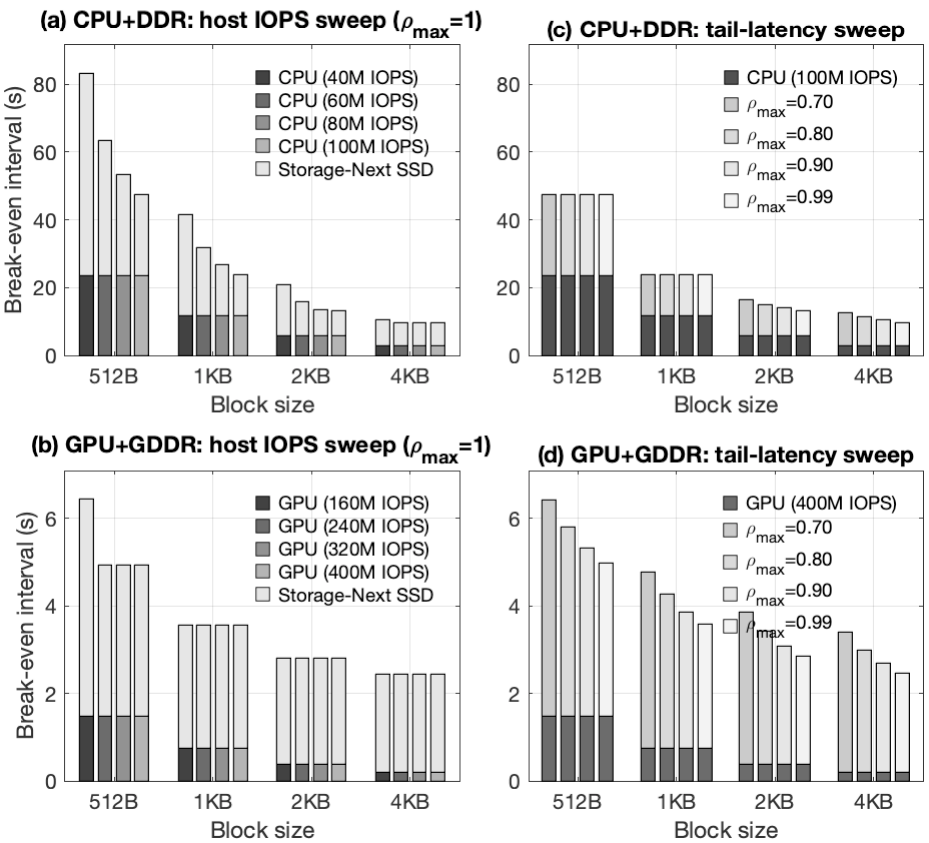}
    \caption{(a) and (b): break-even interval under different host processor IOPS capacity without latency constraint; (c) and (d) break-even interval under different tail latency constraints with fixed processor IOPS capacity.}
    \label{fig:breakevenexample2}
\end{figure}

To model latency constraints, we treat each NAND flash channel as an M/D/1 queue~\cite{kleinrock1975queueing, harchol2013performance}, where read requests follow a Poisson process, service time is deterministic, and one channel serves a single request at a time. Given the peak SSD IOPS \(IOPS_{\text{SSD}}^{\text{(peak)}}\) (see Eq.~\ref{eq:SSDIOPSpeak}) and \(N_{\text{CH}}\) channels, the per-channel service time is \(N_{\text{CH}}/IOPS_{\text{SSD}}^{\text{(peak)}}\). We further include NAND sensing latency and define the channel utilization as \(0 \le \rho \le 1\); the mean read latency is then expressed as
\begin{equation}
    \tau_{\text{mean}}(\rho) = \frac{N_{\text{CH}}}{IOPS_{\text{SSD}}^{\text{(peak)}}}\cdot\frac{\rho}{2(1-\rho)} + \tau_{\text{sense}}\, .\nonumber
\end{equation}
Following Kingman’s heavy-traffic limit~\cite{kingman1961single, harchol2013performance}, the waiting time is well-approximated by an exponential distribution, hence we can approximate the $p$-th percentile tail-latency as  
\begin{align}\label{eq:taureqtail}
\tau_{\text{p}}(\rho) =\frac{N_{\text{CH}}}{IOPS_{\text{SSD}}^{\text{(peak)}}}\cdot\frac{\rho}{2(1-\rho)}\cdot \ln \left(\frac{1}{1-p}\right)+ \tau_{\text{sense}}\, .\nonumber
\end{align}
Let $\{\hat{\tau}_{\text{mean}}, \hat{\tau}_{\text{p}} \}$ denote the application-level constraints on mean and $p$-th percentile tail read latency. Given $\{\hat{\tau}_{\text{mean}}, \hat{\tau}_{\text{p}} \}$, we solve for the largest $\rho\in(0,1)$ (denoted as $\rho_{\max}$) that satisfies $\tau_{\text{mean}}(\rho_{\max})\le \hat{\tau}_{\text{mean}}$ and $\tau_{\text{p}}(\rho_{\max})\le \hat{\tau}_{\text{p}}$. Accordingly, we have that the usable SSD IOPS is $IOPS_{\text{SSD}}=\rho_{\max}\cdot IOPS_{\text{SSD}}^{\text{(peak)}}$. In essence, the scaling factor $\rho_{\max}$ reflects the impact of application-level read latency constraints on the usable SSD IOPS. Moreover, let $IOPS_{\text{proc}}^{\text{(peak)}}$ denote the maximum total IOPS that the host processor can practically sustain, we can further calibrate the usable SSD IOPS as
\begin{equation}\label{eq:rhomaxprocpeak}
   IOPS_{\text{SSD}}= \min \left(\rho_{\max}\cdot IOPS_{\text{SSD}}^{\text{(peak)}}, IOPS_{\text{proc}}^{\text{(peak)}}/N_{\text{SSD}}\right),\nonumber
\end{equation}
where $N_{\text{SSD}}$ is the  number of SSDs. Fig.~\ref{fig:breakevenexample2} extends the quantitative study in Section~\ref{sec:casestudy1} under the feasibility constraints discussed above in this section. We focus on SLC NAND and Storage-Next SSDs (scalable small-block IOPS). Because device service time depends on block size, we specify a separate 99th-percentile read-latency target for each block size, denoted \(\tau_{\text{tail\_512B}}, \tau_{\text{tail\_1KB}}, \tau_{\text{tail\_2KB}}, \tau_{\text{tail\_4KB}}\). For simplicity, we do not set any constraint on mean read latency. Table~\ref{tab:latencyconstraint} gives four tail-latency tiers chosen so that 512B, 1KB, 2KB, and 4KB all admit the same \(\rho_{\max}\in\{0.70,0.80,0.90,0.99\}\). We assume the host drives four SSDs and sweep CPU capacities \(IOPS_{\text{proc}}^{(\mathrm{peak})}\in\{40\text{M}, 60\text{M}, 80\text{M}, 100\text{M}\}\) (guided by \(\sim\!1\)M IOPS/core) and GPU capacities \(IOPS_{\text{proc}}^{(\mathrm{peak})}\in\{160\text{M}, 240\text{M}, 320\text{M}, 400\text{M}\}\) (guided by \(\sim\!4\)M IOPS/SM).
\begin{table}[hbp]
\centering
\caption{99th-percentile tail latency tiers per block size (Storage-Next SSD with SLC NAND), chosen to equalize the admissible utilization \(\rho_{\max}\) across block sizes.}
\vspace*{-6pt}
  \setlength{\tabcolsep}{4pt}
  \renewcommand{\arraystretch}{1.1}
\label{tab:latencyconstraint}
    \begin{tabularx}{0.95\linewidth}
    {|C|C|C|C|C|}
    \hline
         $\tau_{\text{tail\_512B}}$ & $\tau_{\text{tail\_1KB}}$ & $\tau_{\text{tail\_2KB}}$ & $\tau_{\text{tail\_4KB}}$ & 
         $\rho_{\text{max}}$\\
         \hline
         7$\mu s$ & 9$\mu s$ &  11$\mu s$ & 16$\mu s$ & 70\% \\
         \hline
         9$\mu s$ & 11$\mu s$ &  15$\mu s$ & 23$\mu s$ & 80\% \\
         \hline
         13$\mu s$ & 17$\mu s$ &  26$\mu s$ & 44$\mu s$ & 90\% \\
         \hline
         85$\mu s$ & 135$\mu s$ &  230$\mu s$ & 418$\mu s$ & 99\% \\
         \hline
    \end{tabularx}
\end{table}

\paragraph{Impact of host IOPS capacity}
Fig.~\ref{fig:breakevenexample2}(a)-(b) show the effect of the host-side IOPS ceiling $IOPS_{\mathrm{proc}}^{(\mathrm{peak})}$ without latency limits ($\rho_{\max}=1$). In the \emph{host-limited} regime, increasing $IOPS_{\mathrm{proc}}^{(\mathrm{peak})}$ lets more requests be served within the host’s budget, shortening the break-even interval. Once the SSD peak $IOPS_{\mathrm{SSD}}^{(\mathrm{peak})}$ becomes the bottleneck, further increases have no effect. The transition from host- to device-limited behavior depends on both the host budget and block size, since Storage-Next SSD IOPS drop with larger blocks. For example, at 512B on CPU+DDR, raising the CPU budget from 40M to 100M IOPS reduces the interval from 83s to 47s, whereas at 4KB it remains near 10s, indicating a device limitation. GPUs, with far higher $IOPS_{\mathrm{proc}}^{(\mathrm{peak})}$, operate almost entirely in the device-limited regime and, due to better IOPS/\$, sustain shorter intervals, well below 7s across all block sizes.

\paragraph{Impact of latency constraint} Fig.~\ref{fig:breakevenexample2}(c)-(d) hold the host budgets fixed (CPU: 100M IOPS; GPU: 400M IOPS) and vary only the 99th-percentile tail-latency tier from Table~\ref{tab:latencyconstraint}. Tightening the tier (moving from the 99\% row toward 90–70\%) lowers the admissible SSD IOPS utilization \(\rho_{\max}\) and hence usable SSD IOPS, leading to a longer break-even interval. Conversely, when the fixed host budget is already the limiter for a given block size, adjusting the tail tier has little or no effect (e.g., 512B and 1KB on CPU+DDR platform). Quantitatively, the sensitivity to tail latency is modest: for 512B on GPU+GDDR, relaxing the 99th-percentile from 7$\mu$s to 85$\mu$s reduces the break-even interval by only about 1.5s. 

In summary, host processor IOPS capacity is the dominant factor in reducing the break-even interval, whereas latency constraints play a minor role. Increasing the host budget moves the system out of the host-limited regime, lowering the SSD term and producing large, steady gains, especially at small block sizes where devices sustain high IOPS. In contrast, adjusting the tail-latency target changes utilization only slightly. This asymmetry underscores the value of GPUs as I/O engines: their higher IOPS capacity, combined with Storage-Next SSD scalability, consistently drives the break-even interval into the few-seconds regime.

%%%%%%%%%%%%%%%%%%%%%%%%%%%%%%%%
%%%%%%%%%%%%%%%%%%%%%%%%%%%%%%%%
\section{Workload-Aware Platform Analysis (RQ3)} \label{sec:RQ3}\label{sec:actionableprovisioning}
Building on Sections~\ref{sec:RQ1}--\ref{sec:Constraint-aware}, this section introduces a workload-aware framework for quantitatively evaluating a hardware platform’s \emph{viability} and \emph{economic optimality}. Given a workload’s access-interval profile and a fixed platform, we further ask: (i) does the system meet throughput and latency targets, and if so, can it operate at the economics-optimal point? (ii) if not, which hardware resource is the limiting factor, and what upgrade achieves viability or optimality?

%%%%%%%%%%%%%%%%%%%%%%%%%%%%%%%%%%%%%%
\subsection{Analysis Framework Development}\label{sec:viableanalysis}
For simplicity, we assume a single data access granularity \(l_{\text{blk}}\). Let \(N_{\text{blk}}\) be the number of size-\(l_{\text{blk}}\) blocks in the working set (hence total size \(N_{\text{blk}}\cdot l_{\text{blk}}\)). To capture the workload data access characteristics, let \(\tau_i\) denote the average access interval of block \(i\), and define \(\mathcal{S}(T)=\{\,i:\tau_i\le T\,\}\), the set of blocks whose access intervals do not exceed \(T\). The workload also provides mean/tail latency targets and a read:write ratio. A given hardware platform has fixed host-processor IOPS budget, per-SSD peak IOPS, number of SSDs, host-DRAM bandwidth and capacity, and component-cost structure. This work solely focuses on regimes where the working set is much larger than the host-DRAM capacity.

We assume optimal DRAM caching: every block cached in DRAM has a shorter access interval than any uncached block; equivalently, there exists a threshold \(T\) with cached set \(\mathcal{S}(T)\). For any \(T\), the aggregate cached and uncached access throughputs (bytes/s) are
\begin{equation}
\Psi_{c}(T)=l_{\text{blk}}\sum_{i\in\mathcal{S}(T)}\frac{1}{\tau_i},
\qquad
\Psi_{d}(T)=l_{\text{blk}}\sum_{i\notin\mathcal{S}(T)}\frac{1}{\tau_i}.\nonumber
\end{equation}
We assume a zero-copy I/O stack to minimize I/O-induced host-DRAM traffic. Under this model, a DRAM cache miss incurs one SSD\(\rightarrow\)DRAM DMA plus one DRAM read by the processor. The resulting host-DRAM bandwidth demand is
\begin{equation}\label{eq:BDRAM}
B_{\text{DRAM}}^{\text{use}}(T)=\Psi_{c}(T)+2\,\Psi_{d}(T).
\end{equation}
As \(T\) increases, \(\mathcal{S}(T)\) expands, \(\Psi_{c}(T)\) increases, and \(\Psi_{d}(T)\) decreases with \(\Psi_{c}(T)+\Psi_{d}(T)=l_{\text{blk}}\sum_i 1/\tau_i\) fixed; therefore \(B_{\text{DRAM}}^{\text{use}}(T)\) decreases strictly with \(T\). We now define three thresholds, \(T_B\), \(T_S\), and \(T_C\), to isolate the impacts of DRAM bandwidth, SSD bandwidth, and DRAM capacity, respectively. Because \(\mathcal{S}(T)\) expands monotonically with \(T\), the thresholds below are thus well defined and unique whenever a solution exists.

\noindent\textbullet\ \emph{DRAM bandwidth.}
The DRAM-bandwidth threshold \(T_B\) is the \emph{smallest} access-interval threshold at which the required host-DRAM traffic does not exceed the DRAM bandwidth:
\begin{equation}
T_B \;\triangleq\; \min\{\,T>0:\ B_{\text{DRAM}}^{\text{use}}(T)\le B_{\text{DRAM}}\,\},
\end{equation}
where \(B_{\text{DRAM}}\) denotes the host-DRAM bandwidth. Because \(B_{\text{DRAM}}^{\text{use}}(T)\) decreases strictly with \(T\), the solution, when it exists, is unique, and an existence check is \(B_{\text{DRAM}}\ge l_{\text{blk}}\sum_i 1/\tau_i\). %Moreover, increasing \(B_{\text{DRAM}}\) lowers \(T_B\).

\noindent\textbullet\ \emph{SSD bandwidth.}
The SSD-bandwidth threshold \(T_S\) is the \emph{smallest} threshold that confines the uncached throughput to the aggregate usable SSD bandwidth. Given the latency targets and the host-processor IOPS budget, the usable per-SSD \(\mathrm{IOPS}_{\text{SSD}}\) is obtained as in Section~\ref{sec:Constraint-aware}. We define
\begin{equation}
T_S \;\triangleq\; \min\{\,T>0:\ \Psi_d(T)\le B_{\text{SSD}}\,\},\,
\end{equation}
where $B_{\text{SSD}} \;=\; l_{\text{blk}}\cdot N_{\text{SSD}}\cdot \mathrm{IOPS}_{\text{SSD}}$ and \(N_{\text{SSD}}\) is the number of SSDs. Since \(\Psi_d(T)\) decreases with \(T\), the solution is unique. Scaling \(N_{\text{SSD}}\), selecting higher-IOPS devices, or increasing the host-IOPS budget raises \(B_{\text{SSD}}\) and therefore reduces \(T_S\).

\noindent\textbullet\ \emph{DRAM capacity.}
The capacity threshold \(T_C\) is the \emph{largest} threshold whose cached set fits within available DRAM:
\begin{equation}
T_C \;\triangleq\; \max\{\,T>0:\ |\mathcal{S}(T)|\,l_{\text{blk}}\le C_{\text{DRAM}}\,\},
\end{equation}
where \(C_{\text{DRAM}}\) denotes host-DRAM capacity. Ordering \(\{\tau_i\}\) increasingly and letting \(K=\lfloor C_{\text{DRAM}}/l_{\text{blk}}\rfloor\), \(T_C\) equals the \(K\)-th smallest \(\tau_i\); operationally, at most the \(K\) most frequently accessed blocks can be cached in DRAM.

If $\max(T_B, T_S) \le T_C$, the platform is viable for the workload. When $T_B = T_S$, DRAM and SSD bandwidths are balanced. To minimize DRAM cost while maintaining viability, we should select $C_{\text{DRAM}}$ so that $T_C = \max(T_B, T_S)$. The platform operates at the economics-optimal point if $\tau_{\text{break-even}} \in [\max(T_B, T_S),\, T_C]$. If this condition is not met, we should diagnose the limiting path and upgrade accordingly: when $T_B > T_C \ge T_S$, the system is DRAM-bandwidth limited and we should increase $B_{\text{DRAM}}$; when $T_S > T_C \ge T_B$, the storage I/O path is limiting and we should raise the aggregate SSD throughput $B_{\text{SSD}}$, and/or increase the host IOPS budget if $\mathrm{IOPS}_{\text{proc}}^{(\mathrm{peak})}$ is the sub-limiter. When both $T_B$ and $T_S$ exceed $T_C$, bandwidth and capacity are jointly insufficient, so we should either increase $C_{\text{DRAM}}$ until $T_C \ge \max(T_B, T_S)$ or reduce $\max(T_B, T_S)$ through bandwidth upgrades, with the choice guided by a price model or stated priority. After any upgrade, we should recompute $\tau_{\text{break-even}}$; if $\tau_{\text{break-even}} \notin [\max(T_B, T_S),\, T_C]$, the configuration remains viable but off the optimum, and we should further adjust the limiting resources to bring the break-even placement within reach.

%%%%%%%%%%%%%%%%%%%%%%%%%%%%%%%%%%%%%%
\subsection{Quantitative Study}\label{sec:viabilitycasestudy}
Extending the study in Sections~\ref{sec:RQ1}–\ref{sec:Constraint-aware}, we demonstrate the workload-aware platform analysis framework on CPU+DDR and GPU+GDDR platforms. For CPU+DDR platform, we set 12 channels of DDR5-5600 (hence total 540GB/s DRAM bandwidth); for GPU+GDDR platform, we set 8 channels of GDDR6-20 (hence total 640GB/s DRAM bandwidth). We set the peak CPU IOPS capacity to 100M and peak GPU IOPS capacity to 400M. Each platform deploys four SSDs, and we consider both normal SLC SSD and  Storage-Next SLC SSD. We  adopt 99th-percentile read-latency constraint of 13$\mu$s (512B), 17$\mu$s (1KB), 26$\mu$s (2KB), and 44$\mu$s (4KB), corresponding to SSD IOPS utilization $\rho_{\text{max}}$ of 90\% as shown above in Section~\ref{sec:Constraint-aware}. Given the target workload, we aim to make each hardware platform viable (and economics-optimal if practically possible) by provisioning the DRAM capacity $C_{\text{DRAM}}$. Since DRAM capacity now is a variable, we only need to calculate the metrics $T_B$ and $T_S$. Accordingly, we obtain $T_v=\max(T_B, T_S)$ as the viability threshold, and $C_{\text{DRAM}}^{(\text{V})}=|\mathcal{S}(T_v)|\cdot l_{\text{blk}}$ represents the minimum DRAM capacity for making the hardware platform viable. Given the break-even interval $\tau_{\text{break-even}}$, we obtain $T_o=\max(\tau_{\text{break-even}}, T_v)$ as the economics-optimal threshold, and $C_{\text{DRAM}}^{(\text{O})}=|\mathcal{S}(T_o)|\cdot l_{\text{blk}}$ represents the minimum DRAM capacity for making the hardware platform economics-optimum.

\begin{figure}[htbp]
    \centering
    \includegraphics[width=\linewidth]{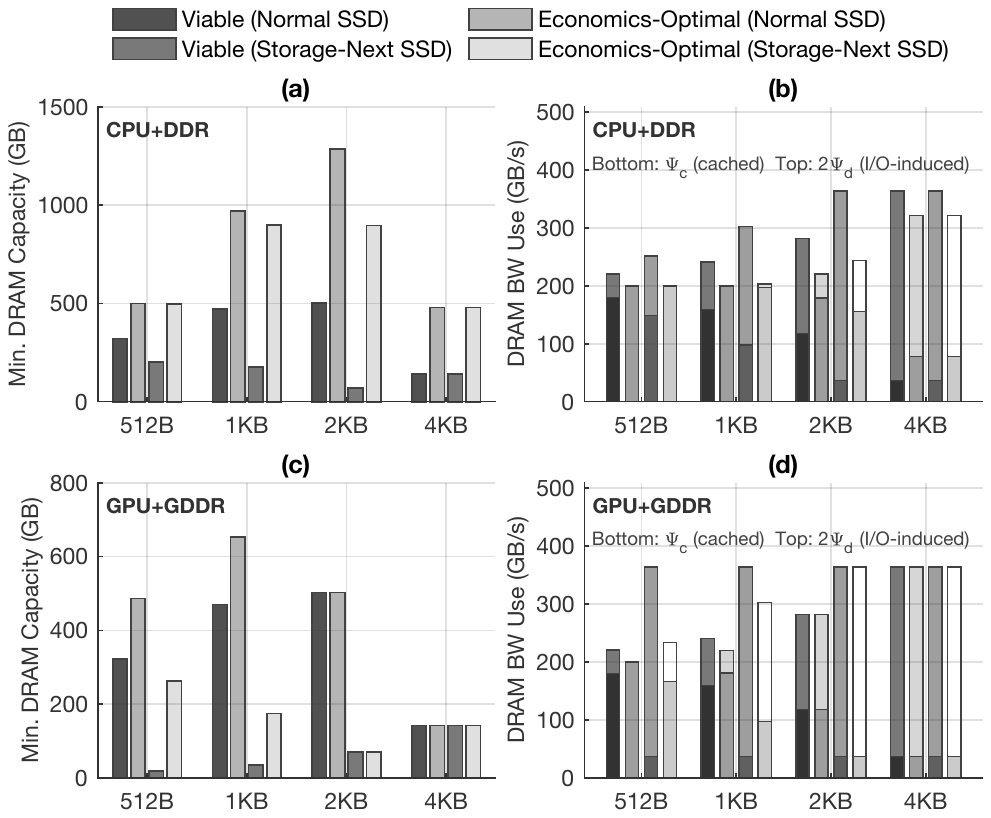}
    \caption{Minimum DRAM capacity required for the CPU+DDR or GPU+GDDR hardware platform to be viable or economics-optimal, and the corresponding DRAM bandwidth usage.}
    \label{fig:casestudy3}
\end{figure}

We assume the workload’s access intervals follow a log-normal distribution with total throughput $l_{\text{blk}}\sum_i 1/\tau_i$ of 200GB/s, comparable to the host DRAM bandwidth. The dataset contains one billion blocks, yielding total sizes of 512GB, 1TB, 2TB, and 4TB for block sizes of 512B, 1KB, 2KB, and 4KB, respectively. Fig.~\ref{fig:casestudy3} shows the minimum DRAM capacities $C_{\text{DRAM}}^{(\text{V})}$ and $C_{\text{DRAM}}^{(\text{O})}$ and corresponding DRAM bandwidth usage. In Fig.~\ref{fig:casestudy3}(b) and (d), each bar shows I/O-induced bandwidth for uncached data (top) and cached data (bottom), corresponding to $2\Psi_d(T)$ and $\Psi_c(T)$ in Eq.~\ref{eq:BDRAM}. Some bars, such as the economics-optimal case under normal SSD at 512B on CPU+DDR, contain only one component because the DRAM already holds the full dataset, eliminating I/O traffic. Overall, as DRAM capacity increases, more data remain resident, misses decline, and I/O-induced bandwidth drops; with less DRAM, a larger uncached set drives more SSD accesses and higher bandwidth demand.

\begin{figure*}[htbp]
    \centering
\includegraphics[width=\linewidth]{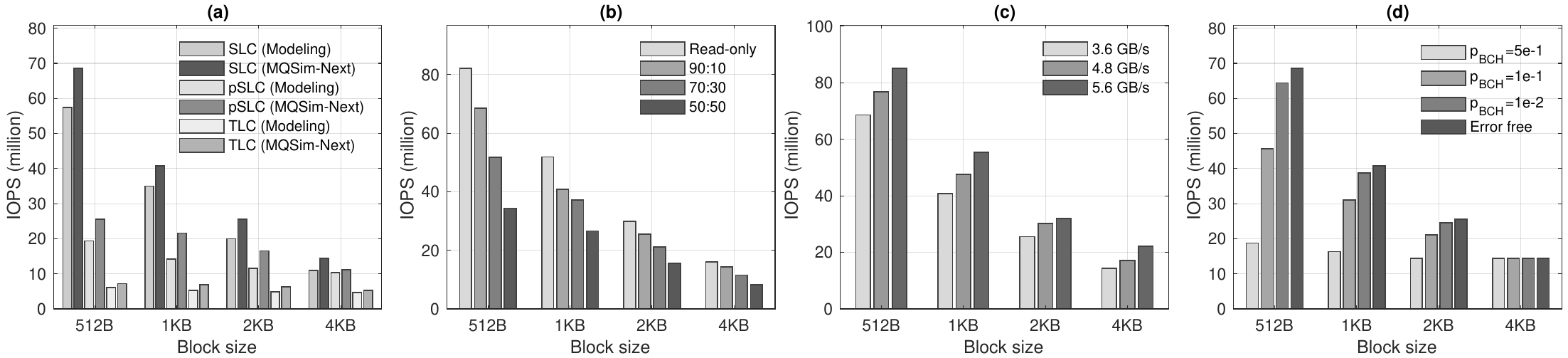}
    \caption{(a) Comparison of modeled and simulated IOPS under 90:10 read-to-write ratios, (b) simulted SLC SSD IOPS under different read-to-write ratios, (c) simulated SLC SSD IOPS under different NAND channel bandwidths, and (d) simulated SLC SSD IOPS under different BCH decoding failure rates, $p_{\text{BCH}}$.}
    \label{fig:MQSimNextResults}
\end{figure*}

On the CPU+DDR platform, whether paired with a normal SSD or a Storage-Next SSD, the break-even interval \(\tau_{\text{be}}\) is consistently longer than \(T_v=\max(T_B,\,T_S)\). Consequently, the economics-optimal DRAM capacity is set by \(\tau_{\text{be}}\), not by viability. At 512B and 1KB block sizes, \(\tau_{\text{be}}\) is so large that achieving the economics optimum requires caching essentially the entire dataset (about 512GB and 1TB, respectively) in DRAM. As block size increases, \(\tau_{\text{be}}\) decreases, so the economics-optimal cache constitutes a smaller fraction of the dataset. Because DRAM bandwidth comfortably exceeds the workload bandwidth, we have \(T_v=T_S\), i.e., SSD IOPS, not DRAM bandwidth, determines the minimum DRAM capacity needed for viability. This explains why the viable DRAM capacity is lower with Storage-Next SSDs: their higher IOPS reduce \(T_S\) and therefore the required cache for viability.

On the GPU+GDDR platform with Storage-Next SSDs, both \(T_B\) and \(T_S\) are small~($<$5s) thanks to high GDDR bandwidth, large GPU IOPS capacity, and high usable SSD IOPS. Consequently, the viable DRAM requirement is low—especially at small block sizes, so the workload can remain viable while a larger share of traffic is serviced as I/O through GDDR. By contrast, the economics-optimal DRAM at 512B and 1KB can be much larger because the break-even interval dominates; the cost-optimal point therefore caches a substantial portion of the working set (e.g., 260GB on GPU+GDDR). At 2KB and 4KB, \(\tau_{\text{be}}\) shortens and \(T_S\) becomes the governing term, so the viable and economics-optimal DRAM capacities coincide. In short, Fig.~\ref{fig:casestudy3} highlights the fundamental advantage of combining GPUs with Storage-Next SSDs: both viability and (often) economics-optimal operation are achievable with far less DRAM than CPU+DDR.

%%%%%%%%%%%%%%%%%%%%%%%%%%%%%%%%%%%%%%%%%%%%%%%%%%%%%%%%%%%%%%
%%%%%%%%%%%%%%%%%%%%%%%%%%%%%%%%%%%%%%%%%%%%%%%%%%%%%%%%%%%%%%
\section{MQSim-Next: Storage-Next SSD Simulator} \label{sec:mqsim}
To study the Storage-Next regime with realistic fidelity, we extend MQSim~\cite{tavakkol2018mqsim, Mqsim-link} into MQSim-Next, preserving its validated foundations (e.g., PCIe/TLP and FTL/cache timing, request-fetch control, and steady-state preconditioning), while modernizing its NAND back-end. Specifically, we incorporate three key upgrades reflecting contemporary device practice: First, we add SCA~\cite{SCA} on the NAND channel, so command/address movement incurs a much shorter per-command cost; this sustains effective channel bandwidth as request sizes shrink and mirrors current device practice. Second, we add the support of \emph{independent multi-plane reads}~\cite{khakifirooz202130, pekny2022independent}, which can much better exploit intra-die parallelism in modern NAND flash. Third, we add the support of \emph{explicit transfer-sense overlap}, allowing array sensing/programming for one request to proceed concurrently with command/address or data movement for another. Together they align MQSim-Next with modern NAND timing and concurrency. The back-end scheduler is correspondingly enhanced with read-prioritized, plane-aware arbitration, allowing short reads to overlap long programs and interleaving SCA bursts before data transfer to maximize small-I/O utilization.

Another key enhancement is an explicit, configurable ECC model. Conventional SSDs protect data in 4KB codewords, flattening random-IOPS for $\le$4KB requests because each small read triggers a full-page decode and transfer. To reach Storage-Next small-block IOPS, MQSim-Next adopts a two-layer concatenated code \cite{macwilliams1977theory,ECC-Shu-book-83}: a BCH inner code per 512B sector and an LDPC outer code spanning eight sectors. Reads touching only a subset of sectors decode the necessary BCH words and skip the LDPC, eliminating intra-SSD read amplification; any BCH failure escalates to a full 4KB LDPC decode that adds transfer and iterative-decode latency. A tunable BCH-error probability parameter allows users to explore small-read tail latency and ECC-induced amplification effects.

We further extend MQSim-Next to support a much larger number of I/O queues, enabling full random-IOPS extraction under deep host parallelism. The simulator is configured using Table~\ref{tab:keySSDpara} parameters and a Gen7~$\times$8 PCIe link\footnote{Chosen to prevent PCIe bandwidth from bottlenecking 4KB IOPS as NAND channel bandwidth scales from 4.8GB/s to 5.6GB/s.}. Fig.~\ref{fig:MQSimNextResults} compares the analytic model and MQSim-Next. The two align closely, with MQSim-Next reporting slightly higher IOPS due to a conservative write-amplification factor ($\Phi_{\mathrm{WA}}\!=\!3$) in the model. Both assume a controller-nonlimiting setup, so observed limits stem from NAND and channel behavior. As shown in Fig.~\ref{fig:MQSimNextResults}(b), IOPS decreases from 82M (read-only) to 68M (90:10), 52M (70:30), and 34M (50:50) as GC traffic competes with host I/O. In Fig.~\ref{fig:MQSimNextResults}(c), increasing channel bandwidth raises IOPS from 68M at 3.6GB/s to 85M at 5.6GB/s, highlighting the benefit of wider channels or stacked I/O such as Xtacking~\cite{huo2022unleash}. Fig.~\ref{fig:MQSimNextResults}(d) shows that 512B BCH failures triggering 4~KB LDPC decodes reduce throughput modestly, remaining near the error-free plateau for $\le$1\% failure rate. Overall, MQSim-Next reproduces the modeled trends~(i.e., workload mix, channel bandwidth, and ECC sensitivity) validating the analytical framework and establishing a reliable foundation for the feasibility and design-space studies that follow.

%%%%%%%%%%%%%%%%%%%%%%%%%%%%%%%%%%%%%%%%%%%%%%%%%%%%%%%%%%%%%%
%%%%%%%%%%%%%%%%%%%%%%%%%%%%%%%%%%%%%%%%%%%%%%%%%%%%%%%%%%%%%%
\section{Re-think Data-intensive Software (RQ4)} \label{sec:RQ4}
By collapsing the DRAM-SSD caching threshold to seconds, Storage-Next SSDs enable a fundamental rethink of algorithm and data-structure design. Two principles guide this shift: (1) re-architect algorithms to exploit ultra-high random IOPS at small block sizes, favoring fine-grained access and concurrency; and (2) leverage flash’s far lower \$/GB than DRAM to use sparse or over-provisioned structures when they improve speed or simplicity, even at higher space cost. Guided by these ideas, this section presents two case studies, i.e., KV~(key-value) store and ANN~(approximate nearest neighbor) search, to demonstrate practical paths to high throughput and simpler, data-intensive software.

\noindent \textbf{Methodology for case studies.} The case-study results in this section are obtained by combining the analytical framework developed in Sections~\ref{sec:RQ1}, \ref{sec:Constraint-aware}, and \ref{sec:RQ3} with device-level characterization from MQSim-Next (Section~\ref{sec:mqsim}). Specifically, MQSim-Next is used to extract peak SSD IOPS and latency behavior under the configurations in Table~\ref{tab:case-ssd}, including block size, read/write ratio, channel bandwidth, and ECC parameters. These device-level characteristics are then incorporated into the analytical feasibility framework to determine usable IOPS under host and tail-latency constraints. Application behavior~(e.g., GET:PUT ratios, reuse-interval distributions, and locality profiles) is modeled using calibrated synthetic traces consistent with the workload assumptions described in each case study. The analytical model determines platform-level break-even and feasibility thresholds, while MQSim-Next provides realistic device behavior to anchor these evaluations.

%%%%%%%%%%%%%%%%%%%%%%%%%%%%%%%%%%%%%%%%%%%%%%%%%%%%%%%%%%%%%%
\begin{figure*}[htbp]
    \centering
\includegraphics[width=\linewidth]{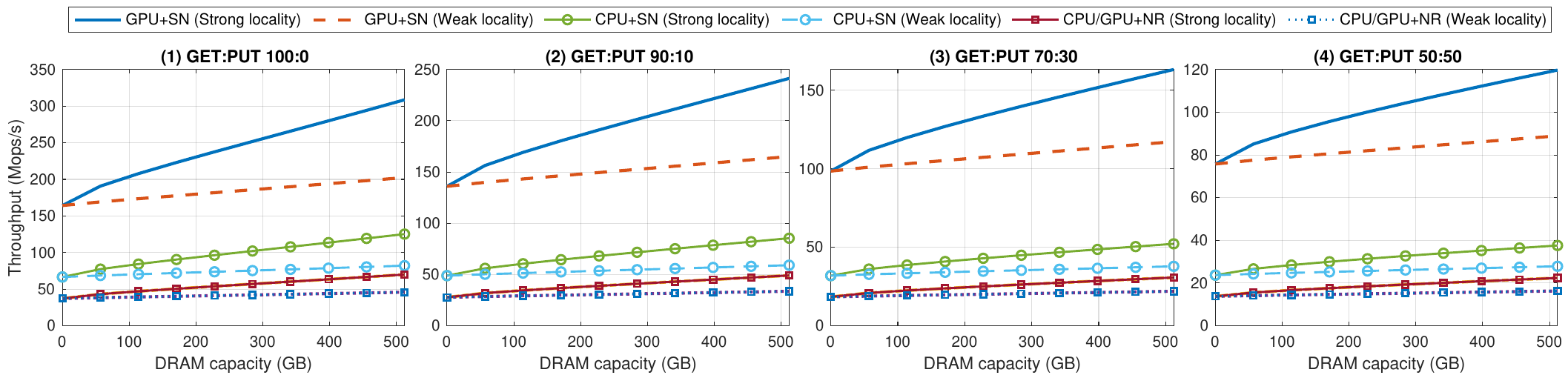}
    \caption{Achievable operational throughput of SSD-resident blocked-Cuckoo KV store under different GET:PUT ratio and DRAM capacity. Storage-Next SSDs and normal SSDs are denoted as SN and NR, respectively. }
    \label{fig:KVstoreres}
\end{figure*}

%%%%%%%%%%%%%%%%%%%%%%%%%%%%%%%%%%%%%%%%%%%%%%%%%%%%%%%%%%%%%%
\subsection{Case Study 1: SSD-Resident KV Store}
\label{sec:KVStorecasestudy}
KV stores anchor modern AI stacks, powering feature lookups in recommenders, embedding caches, LLM memory layers, and session-state in serving pipelines. These workloads often involve billions of unique keys with sparse, unpredictable access patterns. To meet throughput, many systems use in-memory KV stores (e.g., Redis~\cite{Redis-link}, FASTER~\cite{chandramouli2018faster}, MICA~\cite{lim2014mica}) with hash indexing for low latency and simplicity, but the DRAM footprint becomes economically untenable at scale. This has driven interest in hybrid DRAM/SSD engines (e.g., RocksDB~\cite{RocksDB-link,dong2021rocksdb}, WiredTiger~\cite{WiredTiger-link}, Bw-tree~\cite{bwtree-13}) that embrace block I/O and tree-like indexing. However, even these hybrid designs often retain substantial DRAM-resident indexing and metadata (e.g., hash directories, filters, block catalogs) to keep lookup latency acceptable, which grows with key cardinality and still limits capacity per dollar. We distinguish these classical KV stores from LLM KV tensor caches used in attention layers. KV tensor caches typically operate at much coarser granularities (e.g., 32–64KB blocks) and are primarily bandwidth-bound rather than IOPS-bound. In contrast, the workloads considered here emphasize small-block, IOPS-sensitive access patterns common in control-plane state, embedding lookups, long-context reasoning metadata, and agentic execution pipelines. Notably, in the coarse-grained KV tensor regime, the economic break-even interval becomes even shorter than the seconds-scale result derived for small blocks, further strengthening the case for NAND flash as an active tier.

Building on Storage-Next context, we propose an SSD-native KV store that instantiates a blocked Cuckoo hash table~\cite{pagh2004cuckoo,kirsch2010more} on SSDs, eliminating any DRAM-resident index or metadata. Unlike Meta’s hash-based  CacheLib~\cite{berg2020cachelib, CacheLib-link} that discards entries when buckets overflow, we target a persistent KV store that must not drop items, so we adopt Cuckoo hashing, using relocations rather than discards to handle bucket overflows. Each key maps to two candidate SSD-resident buckets, and each bucket matches to one SSD block. Each lookup requires one or two SSD block reads (on average 1.5). To avoid insertion failure, the load factor \(\alpha\) must remain below the critical threshold \(\alpha_{\text{critical}}\) determined by the bucket size \(B=\lfloor l_{\text{blk}}/l_{\text{KV}}\rfloor\), where \(l_{\text{blk}}\) is the SSD block size and \(l_{\text{KV}}\) is the average KV-pair size (e.g., 64B). Prior work~\cite{pagh2004cuckoo,kirsch2010more} shows that even for modest \(B\) (\(\ge 4\)), \(\alpha_{\text{critical}}\) typically exceeds \(0.95\). Insertions may trigger short displacement chains whose expected length can be estimated by
\(
\mathbb{E}[L] \;\approx\; \frac{\alpha^{2B}}{1-\alpha^{B}},
\)
so operating well below \(\alpha_{\text{critical}}\) keeps \(\mathbb{E}[L]\!\ll\!1\), yielding nearly constant insertion latency. We dedicate all available DRAM to caching individual hot KV pairs. We use an SSD-resident write-ahead log (WAL) for persistence and to amortize write cost by consolidating updates that target the same hash bucket. When the WAL exceeds a size threshold, the system commits the consolidated updates into blocked-Cuckoo hash blocks and then recycles the freed log space. 

For demonstration, we evaluate throughput in a realistic large-scale setup: a 5TB KV store with 80 billion 64B items, load factor 0.7, and bucket sizes matched to device class~(512B on Storage-Next SSDs, 4KB on normal SSDs). All DRAM is devoted to caching hot KV pairs. We considered four different GET:PUT ratios (100:0, 90:10, 70:30, and 50:50), with 20\% of PUTs as inserts and the rest updates. Access intervals follow a log-normal  distribution under two locality regimes: strong~($\sigma=1.2$) and weak~($\sigma=0.4$). Hardware matches prior sections: CPU+DDR or GPU+GDDR, where CPU and GPU IOPS capacities are 100M and 400M, and DDR and GDDR bandwidths are 540GB/s and 640GB/s, respectively. Each platform uses four SSDs (either Storage-Next or normal), with SSD bandwidth utilization capped at 70\% to reduce tail latency. Fig.~\ref{fig:KVstoreres} reports simulated achievable throughput under both device/host-IOPS and DRAM-bandwidth bounds. With normal SSDs the system is device-limited, so CPU and GPU collapse into a single curve. Reported throughput includes both DRAM-served cache hits and SSD-served misses; the implied SSD IOPS demand is therefore scaled by the cache-miss fraction. The results show clear dependence on data access locality and GET:PUT ratio: strong locality extracts more value from added DRAM capacity because the cache captures a larger hot set, converts more data accesses into cache hits, and collapses distinct KV pair updates and hence SSD read-modify-write operations per WAL flush. In contrast, as the write share grows, the system issues more read-modify-write operations to SSD, increasing I/O traffic and reducing the operational throughput.

\addtocounter{figure}{1}
\begin{figure*}[htbp]
    \centering
    \includegraphics[width=\linewidth]{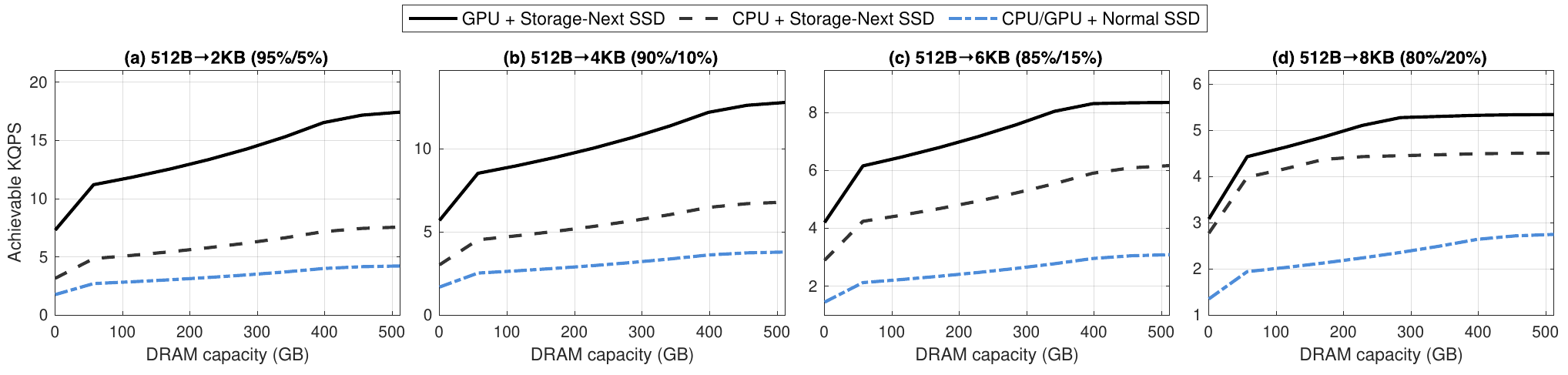}
    \caption{ANN search throughput under different full-vector length with reduced-vector length fixed as 512B.}
    \label{fig:casestudyANN}
\end{figure*}

Pairing GPUs with Storage-Next SSDs (GPU+SN) is especially advantageous. On read-heavy mixes, GPU+SN sustains 100+ Mops/s, comparable to in-memory KV stores such as FASTER~\cite{chandramouli2018faster}. Switching to a CPU with the same Storage-Next SSDs shifts the bottleneck to host IOPS, so throughput falls even though the Storage-Next SSD can deliver more. Across the GET:PUT mixes, DRAM bandwidth becomes the limiting factor only when cache-hit rates are very high; otherwise host IOPS and SSD throughput dominate. In summary, the combination of GPUs \& Storage-Next proves essential for realizing the vision of a fully SSD-resident KV store built on blocked-Cuckoo hashing. By exploiting both the parallelism of GPUs and the IOPS scalability of next-generation SSDs, it transforms NAND flash memory from a passive storage tier into an active, memory-like substrate capable of sustaining in-memory-class KV store throughput.

%%%%%%%%%%%%%%%%%%%%%%%%%%%%%%%%%%%%%%%%%%%%%%%%%%%%%%%%%%%%%%
\subsection{Case Study 2: SSD-Resident ANN Search}
\label{sec:ANNcasestudy}
ANN search is a cornerstone of modern AI services, e.g., recommendation and retrieval-augmented generation (RAG), yet modern workloads often involve TB/PB-scale embedding corpora, well beyond feasible DRAM capacity. Prior SSD-resident systems~\cite{jayaram2019diskann, gollapudi2023filtered} trade search quality to accommodate low IOPS of conventional SSDs.  GPUs paired with Storage-Next SSDs enable a rethink of SSD-resident ANN search. Motivated by widespread use of \emph{dimensionality reduction} in DRAM-resident ANN~\cite{du2016study,deegalla2006reducing,gao2023high,kusupati2022matryoshka}, we propose a \emph{two-stage progressive} SSD-resident design (Fig.~\ref{fig:flow}). Each embedding is stored on SSDs in both a reduced-dimension form (e.g., 512B) and a full-dimension form (e.g., 4KB). At query time, reduced vectors are fetched first to prune unlikely candidates; only a small filtered set is then re-ranked using full vectors. This is effective because most distance computations simply confirm rejection: Gao et al.~\cite{gao2023high} report that over 90\% of comparisons eliminate candidates, so full-dimension evaluation is often unnecessary. Reduced vectors can come from (1) linear transforms such as PCA or random projection~\cite{deegalla2006reducing,weinberger2009distance,du2016study}, (2) a dual-model embedding pipeline, or (3) Matryoshka Representation Learning (MRL)~\cite{kusupati2022matryoshka}, which natively supports multi-resolution vectors. On three MRL-generated corpora (\texttt{MS~MARCO}, \texttt{20~Newsgroups}, \texttt{DBpedia}), our experiments show the progressive scheme sustains recall $>$98\%.
\addtocounter{figure}{-2}
\begin{figure}[htbp]
	\centering
    \includegraphics[width=0.9\linewidth]{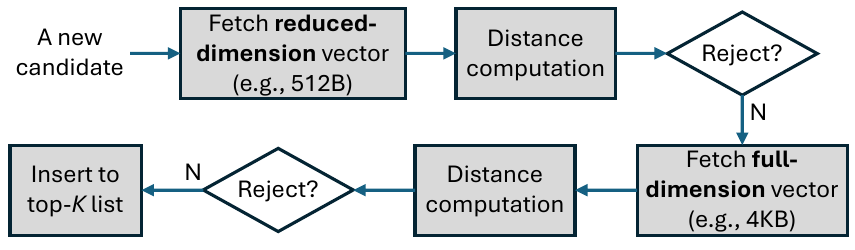}
	\caption{Illustration of two-stage progressive ANN search.}
\label{fig:flow}
\end{figure}
\addtocounter{figure}{1}

The two-stage scheme benefits directly from Storage-Next SSDs: because most accesses hit reduced-dimension vectors, the workload issues predominantly small-block random reads (e.g., 512B), which unlocks very high IOPS and lifts end-to-end throughput. For demonstration, consider an 8\,billion-embedding corpus with full-dimension sizes of 2KB, 4KB, 6KB, and 8KB, respectively, while fixing the reduced dimension at 512B. We focus on HNSW (Hierarchical Navigable Small World)~\cite{malkov2018efficient}, widely used in state-of-the-art ANN search. To improve scalability, we co-locate graph-link metadata with each node on SSD and devote available DRAM to caching the hotter upper-layer nodes. HNSW concentrates traversal by layer: layer sizes shrink rapidly with height, per-query visits per layer also decline (though more slowly), and thus per-node access intervals shorten at higher layers, making them DRAM-cache friendly. We evaluate using a calibrated, layer-aware synthetic trace that mirrors HNSW’s coarse-to-fine pipeline. Under full-dimension sizes of 2KB, 4KB, 6KB, and 8KB, roughly 5\%, 10\%, 15\%, and 20\% of candidates are promoted to full-vector re-ranking, respectively. In this regime, reduced-vector fetches are IOPS-bound and benefit most from Storage-Next SSDs, while the small promoted fraction is bandwidth-bound yet amortized by the large rejection rate.

Fig.~\ref{fig:casestudyANN} shows simulated ANN search throughput (KQPS) versus DRAM capacity under four reduced$\rightarrow$full vector configurations. We use the same GPU+GDDR and CPU+DDR platforms as before, each with four SSDs. Across all scenarios, GPU with Storage-Next SSDs achieves the highest KQPS. In lighter-promotion cases, i.e., (a) 512B$\rightarrow$2KB (95\%/5\%) and (b) 512B$\rightarrow$4KB (90\%/10\%), the GPU setup remains SSD-IOPS-limited, rising from 7–11 KQPS at small DRAM to 13–17 KQPS at 512GB as caching reduces SSD reads. CPU+Storage-Next lies below, capped by the CPU’s 100M IOPS budget. The Normal SSD baseline is always SSD-limited. As the promotion rate increases, DRAM traffic grows and bandwidth ceilings appear. In (c) 512B$\rightarrow$6KB (85\%/15\%), GPU+Storage-Next becomes DRAM-bandwidth-limited beyond 400GB (plateauing near 8.3 KQPS), while the CPU remains host-IOPS-limited (up to 6.2 KQPS). In the heaviest mix (d) 512B$\rightarrow$8KiB (80\%/20\%), GPU+Storage-Next hits the DRAM-bandwidth limit near 300GB, and the CPU transitions from mixed to fully bandwidth-limited beyond 200GB. Overall, Storage-Next SSDs deliver a consistent 2–3× throughput advantage over Normal SSDs: small-block IOPS dominate at low caches, GDDR bandwidth sets the high-cache plateau, and host IOPS capacity determines how much of the SSD’s potential is realized.

Overall, these results show that pairing GPU hosts with Storage-Next SSDs makes low-cost fully SSD-resident ANN practical. TB/PB-scale embedding tables can remain on flash while sustaining high recall and high KQPS, avoiding the large DRAM capacity required for in-memory retrieval. For context, DiskANN~\cite{jayaram2019diskann}, an SSD-resident system from Microsoft that constructs a pruned on-disk graph and streams neighbor lists, achieves roughly 5 KQPS on billion-scale benchmarks. On our modeled hardware, the GPU+Storage-Next configuration pushes this boundary toward tens of KQPS\footnote{Our results are illustrative and model-based, not a direct performance comparison with DiskANN.}. This indicates that the memory hierarchy in the Storage-Next era can match or exceed DiskANN-class throughput while retaining HNSW-level search quality.

%%%%%%%%%%%%%%%%%%%%%%%%%%%%%%%%%%%%%%%%%%%%%%%%%%%%%%%%%%%%%%
%%%%%%%%%%%%%%%%%%%%%%%%%%%%%%%%%%%%%%%%%%%%%%%%%%%%%%%%%%%%%%
\section{Limitations and Future Work}
While this study establishes a first-principles, feasibility-aware framework for the memory-hierarchy paradigm shift enabled by Storage-Next, several simplifying assumptions and open research directions remain.

\noindent\textbf{Device and cost modeling.} Our analysis uses normalized cost parameters and NAND timing values representative of mature 2025-era technologies. Although process variations and future scaling nodes may change absolute numbers, the relative trade-offs, particularly the IOPS-driven collapse of the break-even interval, remain robust. Future work could integrate process-scaling models, cost-learning curves, and the implications of die stacking or 3D integration to capture the economic trajectory of next-generation NAND and controllers more faithfully. Beyond capital cost, one can incorporate operational cost by extending DRAM ``rent'' to include power consumption and extending the per-I/O SSD cost to include dynamic energy per request. This yields a total-cost-of-ownership~(TCO) formulation that captures both CapEx and OpEx.

\noindent\textbf{Endurance and write economics.} We model write amplification in both the analytic and simulation frameworks, but do not yet incorporate device endurance limits (e.g., retention, refresh policies) or lifetime-driven costs. Extending the framework with endurance-aware models, covering lifetime derating, refresh-induced bandwidth taxes, and energy per I/O, would elevate it to a deployment-grade, sustainability-aware provisioning tool.

\noindent\textbf{Workload coverage.} Our workloads focus on read-dominant, large-footprint AI and analytics under single-tenant settings. Extending to write-intensive, transactional, or multi-tenant environments will require modeling of time-varying garbage collection, compaction interference, and bursty access patterns that inflate tail latency. Factoring in update locality and write shaping will improve realism and make the viability analysis broadly applicable across storage services and data-center workloads.

\noindent\textbf{System integration and topology.} The framework assumes optimized local PCIe/NVMe paths and single-node coherence. Future deployments increasingly rely on multi-socket servers and disaggregated fabrics that introduce additional latency domains and queue hierarchies. Extending the analysis to these distributed or composable-memory environments will clarify how seconds-scale caching interacts with remote access and networked storage layers. Future systems increasingly deploy fabric-attached storage (e.g., NVMe-over-Fabrics) and intermediate memory tiers such as CXL-attached memory. The same first-principles break-even formulation can be extended by introducing fabric latency/bandwidth terms and applying the analysis pairwise across adjacent tiers.

\noindent\textbf{Host-side I/O optimization.} Because host IOPS capacity strongly governs the seconds-scale regime, a key direction is to reduce software overheads and co-design host-device interfaces. Promising approaches include: (i) streamlining the I/O stack for reduced submission latency, and (ii) developing lightweight I/O accelerators for queue management and completion coalescing. These efforts point to IOPS-scalable architectures where software and hardware evolve jointly.

\noindent\textbf{Algorithmic design space exploration.} The collapse of caching threshold to seconds blurs the traditional boundary between memory and storage, opening a broad design space for SSD-resident algorithms and data structures that treat flash as an active tier. Rather than prescribing specific mechanisms, we emphasize the scope: access-path design and scheduling at high IOPS; tier-aware data layouts and placement; lightweight ordering, consistency, and recovery tuned to seconds-scale reuse; and QoS, fairness, and isolation under multi-tenant contention. Exploring this spectrum through cross-layer co-design can yield a new class of SSD-resident systems purpose-built for the seconds-scale regime.

%%%%%%%%%%%%%%%%%%%%%%%%%%%%%%%%%%%%%%%%%%%%%%%%%%%%%%%%%%%%%%
%%%%%%%%%%%%%%%%%%%%%%%%%%%%%%%%%%%%%%%%%%%%%%%%%%%%%%%%%%%%%%
\section{Conclusion}
This work re-examines the five-minute rule from first principles and recasts it as a feasibility-aware provisioning framework for AI-era systems. Our analysis shows that, when GPU-centric hosts are paired with Storage-Next SSDs engineered for fine-grained random access, the DRAM-to-flash caching threshold collapses from minutes to seconds, effectively promoting NAND flash to an active extension tier of DRAM. We implemented MQSim-Next to reproduce the key trends and support basic sensitivity studies. We further examined SSD-resident KV store and ANN search as concrete case studies, illustrating the algorithm and data-structure design space unlocked by such a paradigm shift. Overall, this work turns a heuristic into a quantitative, cross-layer framework for the AI era, laying a foundation for treating flash as a first-class citizen in the memory hierarchy, bridging device physics, system design, and algorithm co-optimization.

%%%%%%%%%%%%%%%%%%%%%%%%%%%%%%%%%%%%%%%%%%%%%%%%%%%%%%%%%%%%%%
\section*{Acknowledgments}
The authors gratefully acknowledge the broader NVIDIA Storage-Next community for shaping both the motivation and direction of this work. In particular, we thank our collaborators and industry partners---including NVIDIA, Micron Technology, KIOXIA, SK hynix, Samsung, and H3 Platform---for their deep technical engagement, early evaluations, and candid feedback on emerging storage-system behavior at scale. Their insights into device-level characteristics, firmware dynamics, and telemetry-informed optimization were instrumental in refining our understanding of modern NVMe systems under extreme workloads.

\bibliographystyle{IEEEtranS}
\bibliography{reference} 

%\vfill

\end{document}